\documentclass[aps,prc,twocolumn,showpacs,superscriptaddress]{revtex4-1}
\usepackage{natbib}
\usepackage{graphicx}
\usepackage{color}
\usepackage{slashbox}
\begin{document}
\bibliographystyle{revtex}
\title{Microscopic multiphonon approach to  nuclei with a valence hole in the oxygen region}
\vspace{0.5cm}
\author{G. De Gregorio}
\affiliation{Nuclear Physics Institute,
Czech Academy of Sciences, 250 68 \v Re\v z, Czech Republic} 
\affiliation{INFN Sezione di Napoli, 80126 Napoli, Italy}
\author{F. Knapp}   
\affiliation{Faculty of Mathematics and Physics,  Charles University, 116 36 Prague, Czech Republic } 
\author{N. Lo Iudice}
\affiliation{INFN Sezione di Napoli, 80126 Napoli, Italy}
\affiliation{Dipartimento di Fisica, 
Universit$\grave{a}$ di Napoli Federico II, 80126 Napoli, Italy} 
\author{P. Vesel\'y} 
\affiliation{Nuclear Physics Institute,
Czech Academy of Sciences, 250 68 \v Re\v z, Czech Republic} 
\date{\today}
 
\begin{abstract}
An equation of motion phonon method, developed for even nuclei and recently extended to odd systems with a valence particle,  is formulated in the hole-phonon coupling scheme and applied to A=15 and A=21 isobars with a valence hole. The method derives   a set of equations which yield an orthonormal basis of states composed of a hole coupled to an orthonormal basis of  correlated $n$-phonon  states ($n= 0,1,2,\dots$), built of constituent Tamm-Dancoff phonons, describing the excitations of a doubly magic core. The basis is then adopted to solve the full eigenvalue problem. The method is formally exact but lends itself naturally  to simplifying approximations. Self-consistent calculations using a  chiral Hamiltonian in a space encompassing up to two-phonon and three-phonon basis states  in A=21  A=15 nuclei, respectively,  yield full spectra, moments, electromagnetic and $\beta$-decay transition strengths, and  electric dipole cross sections. The analysis of the hole-phonon composition of the eigenfunctions  contributes to clarify the mechanism of excitation of levels and resonances and to understand the reasons of the deviations of the theory from the experiments. Prescriptions for reducing these discrepancies are suggested.
\end{abstract}
\maketitle

\section{Introduction}

We are witnessing a renewed interest toward the spectroscopic studies of odd nuclei.
Several investigations on heavy nuclei were carried out within the   
particle-vibration coupling (PVC) model using the random-phase-approximation (RPA) or its extension to describe the core excitations.   
Most PVC approaches have exploited   energy density functionals (EDF)  derived  from Skyrme forces \cite{Mizu12,Cao14,Tarpa14} or from relativistic meson-nucleon Lagrangians   \cite{AfaLitv15}
or  based on the theory of finite Fermi systems \cite{Gnez14}.   
Other calculations  adopted the quasiparticle phonon model (QPM) using a separable interaction \cite{MishVor08}
or  a perturbative approach using the Gogny potential \cite{Co15} or were framed within   the interacting boson fermion
model (IBFM) with parameters evaluated microscopically \cite{Nomura17}.

Light odd nuclei were studied within
an equation of motion method  based on the coupled cluster (CC) theory  \cite{Gour06,Hagen10,Hagen12,Jansen14,Hagen14}, a  self-consistent Green's function theory approach \cite{Cipol13}, a no-core shell model (NCSM) \cite{Barrett13},
and  a   many-body perturbation theory calculation \cite{Holt13}. All these investigations adopted  NN + 3N chiral forces derived from effective field theories and were focused  mainly on the bulk properties and low-lying spectra of odd nuclei around $^{16}$O.   
 
Describing the full spectra in this region is quite challenging due to  the complex structure of  $^{16}$O which affects deeply the spectroscopic properties of all surrounding odd nuclei.  We have attempted such a study by adopting the equation of motion phonon method (EMPM).
An orthonormal basis of    $n$-phonon states ($n=1,2, \dots  )$,  built of phonons obtained in  Tamm-Dancoff approximation (TDA), is generated by an appropriate set of equations and adopted for solving the full eigenvalue problem.
This method can be considered an upgrading  of the mentioned microscopic PVC. It includes, in fact, multiphonon states with   an arbitrary number of phonons, takes the Pauli principle into full account, and does not rely on any approximation. It has, in fact, the same accuracy of shell model.

It was first devised for even-even closed shell  \cite{AndLo,AndLo1,Bianco}   and  adopted to investigate the dipole response in   heavy, neutron rich, nuclei   \cite{bianco12,Knapp14,Knapp15}.  It was, then, reformulated in terms of Hartree-Fock-Bogoliubov (HFB)  quasiparticles and employed to study the full spectrum and  the dipole response of the neutron rich open shell $^{20}$O \cite{DeGreg16}.

A particle-phonon version was developed recently and adopted to investigate thoroughly the spectroscopic properties of $^{17}$O and $^{17}$F \cite{DeGreg16a,DeGreg17a,DeGreg17b} as well as the neutron rich $^{23}$O and $^{23}$F \cite{DeGreg18}.

Here, we reformulate the EMPM in the hole-phonon scheme  to investigate $^{15}$O and $^{15}$N and the neutron rich $^{21}$O and $^{21}$N.

The strong impact of many particle-hole (p-h) core excitations on  $^{15}$O and $^{15}$N was ascertained soon after the first excited 0$^+$ state at 6.06 MeV in $^{16}$O was assigned a dominant 4p-4h structure \cite{BrownGreen66}. 
  
In an earlier work \cite{HalFrench57} Halbert and French succeeded in explaining a fraction of the low-lying positive parity states in $^{15}$N
only after the inclusion of leading 1p-2h configurations.

For a more complete description, however, it was necessary to add 3p-4h states  \cite{Shukla68,Lie70,Alburger79,Raman94}. These configurations
could also describe a considerable number of negative parity levels \cite{LieEng76}.

Nuclei around $^{16}$O were investigated in more recent papers. A shell model calculation in a $(p,sd)$ configuration space used the empirical WBM interaction 
\cite{Utsuno2011}. The same interaction and code were adopted for 
$^{15}$N \cite{Mertin2015}.
Another shell model study was focused on  pygmy (P) and giant (G) dipole resonances (DR) in a chain of N isotopes including $^{15}$N \cite{Ma2016}.

Several experiments supported by theoretical analyses based on shell model calculations using empirical forces have been devoted to $^{15}$N \cite{Alburger79,Raman94} as well as to $^{21}$O and $^{21}$N 
\cite{Catford89,Sauvan00,Sauvan04,Stanoiu04,Mueller90,Li09,Zhang16}.

In our EMPM approach, we adopt a HF basis derived from the  chiral NN potential  (NNLO$_{opt}$) optimized so as to minimize the contribution of the three-body term \cite{Ekstr13}. 
This potential, while producing too much attraction in medium and heavy mass nuclei, reproduces well the experimental binding energies of light nuclei and  oxygen isotopes. 

Upon solving the equations of motion, we produce a basis of  states composed of a valence hole coupled to a full set of TDA phonons generated in a large configuration space plus a subset of  two-phonon and, for A=15, three-phonon states, the latter obtained by  an approximate procedure which will be described later.  
The availability of  all eigenvalues and eigenstates allowed by the space dimensions enables us to produce the complete level schemes as well as all  moments and transition strengths, and  to induce the damping and fragmentation  of the GDR and PDR. The phonon composition of the states sheds light on   the excitation mechanisms,  the nature of levels and resonances, and 
provides useful hints for removing the discrepancies between theory and experiments.

\section{EMPM for nuclei with a valence hole}

\subsection{Generation of the core multiphonon basis} 
The   primary goal of the method \cite{Bianco} is to  generate an orthonormal  basis of $n$-phonon correlated states      
\begin{equation}
\mid \alpha_n \rangle = \sum_{ \lambda \alpha_{n-1}} 
C_{\lambda \alpha_{n-1} }^{\alpha_n } \mid (\lambda \times \alpha_{n-1})^{\alpha_n} \rangle
\label{nstatec}
\end{equation}
of energy $E_{\alpha_n}$, where
\begin{equation}
\mid (\lambda \times \alpha_{n-1})^{\alpha_n} \rangle
= \Bigl\{O^\dagger_\lambda \times \mid \alpha_{n-1} \rangle \Bigr\}^{\alpha_n},  
\end{equation}
and
\begin{equation}
O^\dagger_\lambda = \sum_{ph} c^\lambda_{ph} (a^\dagger_p  \times b_h)^\lambda 
\label{Olam}
\end{equation}
is the p-h TDA phonon operator of energy $E_\lambda$ acting on the $(n-1)$-phonon basis states $ \mid \alpha_{n-1}>$, assumed to be known.  The operators $a^\dagger_p=  a^\dagger_{x_p j_p m_p} $  and $b_h = (-)^{j_h + m_h} a_{x_h j_h - m_h}$ create  a particle 
and a hole  of energies $\epsilon_p$ and $ - \epsilon_h$, respectively. 

To this purpose,  we start with  the equations of motion
 \begin{equation}
\label{Eqmo}
 \langle \beta \parallel [H,O^\dagger_\lambda]   \parallel \alpha \rangle  = 
\Bigl(E_{\beta} - E_{\alpha} \Bigr) \langle \beta \parallel O^\dagger_\lambda  \parallel  \alpha \rangle 
\end{equation}
where $\beta$ and $\alpha$ stand for $\alpha_n$ and $\alpha_{n-1}$.  
By making use of  Eq. (\ref{nstatec}), it is possible to express the amplitudes $\langle \beta \parallel O^\dagger_\lambda  \parallel  \alpha \rangle$  in terms of the  expansion coefficients $C_{\lambda \alpha }^{\beta }$ 
\begin{equation}
 \langle \beta \parallel O^\dagger_\lambda  \parallel  \alpha \rangle 
 =   [\beta]^{1/2}\sum_{\lambda' \alpha'} {\cal D}^{\beta}_{\lambda \alpha \lambda' \alpha'} 
C_{\lambda' \alpha' }^{\beta}, 
  \label{XC}
\end{equation}
where $[\beta] = 2 J_{\beta} + 1$, a notation which will be used throughout the paper, and
\begin{equation}
\label{Dev}
 {\cal D}^{\beta}_{\lambda \alpha \lambda' \alpha'}= \langle (\lambda' \times \alpha')^{\beta} \mid (\lambda \times \alpha)^{\beta} \rangle
\end{equation}
is the overlap or metric matrix which reintroduces the exchange terms among different phonons and, therefore, re-establishes   the Pauli principle.    
 
We proceed by expanding   the commutator in Eq. (\ref{Eqmo}) and    expressing the p-h  operators  in terms of the phonon operators $O^\dagger_{\lambda}$ upon inversion of Eq. (\ref{Olam}). We then  exploit Eq. (\ref{XC}) and obtain the generalized eigenvalue equation   
\begin{eqnarray}
      \sum_{\lambda' \alpha' \lambda" \alpha"}  \Bigl( (E_\lambda + E_{\alpha} - E_{\beta} ) \delta_{\lambda \lambda'} 
			\delta_{\alpha \alpha'} 
			+{\mathcal V}^{\beta}_{\lambda \alpha \lambda' \alpha'}\Bigr)\nonumber\\
			\times {\cal D}^{\beta}_{\lambda' \alpha' \lambda" \alpha"} C^{\beta}_{\lambda" \alpha"}
			=0.
 \label{Eig}                                       
\end{eqnarray} 
The formulas giving the metric matrix ${\cal D}^{\beta}_{\lambda \alpha \lambda' \alpha'}$ and the phonon-phonon potential ${\mathcal V}^{\beta}_{\lambda \alpha \lambda' \alpha'}$  can be found in  \cite{Bianco}.

The above eigenvalue equation is singular since the basis $\mid (\lambda \times \alpha)^{\beta} \rangle$ is over-complete. 
Following  the procedure outlined in Refs. \cite{AndLo,AndLo1}, based on the Cholesky decomposition method,
it is possible to extract a basis of linearly independent states 
spanning the physical subspace and obtain a non singular eigenvalue equation whose solution
 yields a basis of orthonormal correlated $n$-phonon states of the form (\ref{nstatec}).  
 
Since recursive formulas hold for all quantities, it is possible to solve the eigenvalue equations    iteratively starting from the TDA phonons 
$\mid \alpha_1 \rangle= \mid \lambda \rangle$ and, thereby, generate a set of orthonormal multiphonon states $\{\mid 0 \rangle, \mid \alpha_1 \rangle   ,  \dots \mid \alpha_n, \rangle  \dots \}$.


\subsection{Eigenvalue problem in the hole-phonon scheme}

For an odd nucleus with a valence hole  we intend to
generate a basis  of hole-core states $\mid \nu \rangle$ of  spin $v$ and energy $E_{\nu}$  having the form   
\begin{equation}
\mid \nu \rangle = \sum_{ h \alpha} C_{h \alpha }^{\nu } \mid (h^{-1} \times \alpha)^v \rangle =
\sum_{ h \alpha} C_{h \alpha }^{\nu } \Bigl\{b_h \times \mid \alpha \rangle \Bigr\}^v,  
\label{nu}
\end{equation}
where $\mid \alpha \rangle$ are  $n$-phonon states of the form (\ref{nstatec}) describing the excitations of the core. 
 
We mimic the procedure adopted for the particle-phonon scheme \cite{DeGreg16a} and 
start with the equations 
\begin{equation}
\langle  \alpha \parallel [b_h,   H]^h \parallel \nu  \rangle  = 
 (E_{\nu} - E_{\alpha}) X^{\nu}_{h \alpha} ,
\label{comeig} 
\end{equation}
where  
\begin{equation}
 X^{\nu}_{h \alpha} = \langle \alpha \parallel b_{h} \parallel \nu \rangle=
[v]^{1/2}\sum_{h' \alpha' } {\cal D}^{v}_ {h \alpha h' \alpha'}
C_{h' \alpha' }^{\nu}
\label{Xpnu}
\end{equation}
and the overlap matrix is given by
\begin{eqnarray}
&&{\cal D}^{v}_ {h \alpha h' \alpha'} = \langle (h^{-1} \times \alpha)^v \mid  (h'^{-1} \times \alpha')^v \rangle=
\delta_{hh'} \delta_{\alpha \alpha'} \nonumber\\
&& + \sum_\sigma [\sigma]^{1/2}   W(\sigma h \alpha' v;  h'  \alpha) \langle \alpha' \parallel 
(a^\dagger_{h'}\times b_h)^\sigma \parallel \alpha \rangle,
\end{eqnarray}
where $W(\sigma h \alpha' v;  h'  \alpha)$ are Racah coefficients.
The second piece reintroduces the exchange terms among the odd hole and the $n$-phonon states and, thereby, re-establishes   the Pauli principle.  

A procedure analogous to the one adopted for even nuclei leads to the generalized eigenvalue equation  
\begin{eqnarray}
\sum_{h' \alpha' h" \alpha"} \Bigl\{ (\epsilon_h + E_\alpha - E_\nu) \delta_{hh'} \delta_{\alpha \alpha'} 
+ {\cal V}^v_{h \alpha h' \alpha'} \Bigr\}\nonumber\\
 \times {\cal D}^v (h' \alpha', h" \alpha") C^{\nu}_{h" \alpha"} =0.
\label{eigChnew}
\end{eqnarray}
${\cal V}^{v}_{h \alpha h' \alpha'}$ is the hole-phonon potential 
 given by
\begin{equation}
{\cal V}^v_{h \alpha h' \alpha'} =  \sum_{\sigma} [\sigma]^{1/2} (-)^{h + h' - \sigma} W(\alpha \sigma v h'; \alpha' h) 
{\cal F}^\sigma_{h \alpha h' \alpha'} 
\end{equation}
where
\begin{eqnarray}
{\cal F}^\sigma_{h \alpha h' \alpha'} =  
\sum_{tq} F^\sigma_{h h'tq}
\langle \alpha \parallel (a^\dagger_t \times b_q )^\sigma \parallel \alpha' \rangle. 
\end{eqnarray}
Here the sum runs over particles $tq= p_1 p_2$ and holes $tq = h_1 h_2$
and $F^\sigma$ is related to the nucleon-nucleon potential $V^\Omega$ by
 the Pandya transformation
\begin{equation}
F^\sigma_{rsqt}  =    \sum_{\Omega}  [\Omega]  (-)^{r+t - \sigma -\Omega} W(rsqt;\sigma \Omega)
  V^\Omega_{rqst}. 
\end{equation}
 Following the  procedure  based on the Cholesky decomposition method adopted in the particle-phonon scheme \cite{DeGreg16a}, we  extract from the  over-complete set  $ \mid (h^{-1} \times \alpha_n)^\nu \rangle$ a basis of 
linearly independent states  and obtain a non singular eigenvalue equation. Its  iterative solution, starting from $n=1$, yields the correlated hole-core states $\mid \nu_n \rangle$ (\ref{nu}) of energies $E_{\nu_n}$ for $n=1,2....$, which, together with the single hole states $\mid \nu_0 \rangle$, form an orthonormal basis.

We have now all the ingredients necessary for solving  the eigenvalue problem in the full space spanned by $\{\mid \nu_0 \rangle, \mid  \nu_1 \rangle, \dots |\nu_n>, \dots \}$   
\begin{equation}            
 \sum_{\nu'_{n'}} \Bigl\{ \big(E_{\nu_n} - {\cal E}_{\nu}  \bigr)  \delta_{\nu_n \nu'_{n'}}+
{\cal V}^{\nu}_{\nu_n  \nu'_{n'}} \Bigr\}{\cal C}^{\nu}_{\nu'_{n'}} =0,
\label{eigfull} 
\end{equation}
where the matrix element of ${\cal V}^{\nu}_{\nu_n  \nu'_{n'}}$ are non vanishing only  for $n' =n +1$  and $n'= n + 2$.

  For $n' = n+1$ we have
\begin{equation}
{\cal V}^{(\nu)}_{\nu_n  \nu'_{n'}} = \frac{1}{[v]^{1/2}}
\sum_{h\alpha_n h' \alpha'_{n'}} C^{(\nu_n)}_{h \alpha_n} {\cal V}^{v}_{h \alpha_n h' \alpha'_{n'}} 
\langle \alpha'_{n'} \parallel a^\dagger_{h'} \parallel \nu'_{n'} \rangle,
\label{Vnnp}
\end{equation}
where
\begin{eqnarray}
\label{Vhahpap}
&&{\cal V}^{\nu}_{h \alpha_n h' \alpha'_{n'}} = - \delta_{h h'}  
\langle \alpha_n \mid H \mid \alpha'_{n'} \rangle + \nonumber\\ 
&& \sum_{\lambda} [\lambda]^{1/2}  
{\cal F}^{\lambda}_{h h'}
W(\alpha_n \lambda v h'; \alpha'_{n'} h), 
\end{eqnarray} 
and
\begin{eqnarray}
{\cal F}^{\lambda}_{h h'} = 
\sum_{ h_1 p_1 } F^{\lambda}_{h h' p_1 h_1} c^{\lambda}_{p_1 h_1}.
\end{eqnarray}
 For $n'=n+2$, we have simply
\begin{equation}
\label{Vnn2}
{\cal V}^v_{\nu_n \nu'_{n'}}   
=   \sum_{\alpha_2} \langle \alpha_2 \mid H \mid 0 \rangle \langle \nu'_{n'} \mid (\nu_n \times \alpha_2)^v \rangle.
\end{equation}
Eq. (\ref{eigfull}) yields all eigenvalues and eigenstates allowed by the dimensions of the multiphonon space.
The  eigenfunctions have the composite structure
\begin{equation}
\mid \psi_\nu \rangle = \sum_{\nu_n} {\cal C}^\nu_{\nu_n} \mid \nu_n \rangle,
\label{Psi}
\end{equation}
where $\mid \nu_n \rangle$ are given by Eq. (\ref{nu}).

\subsection{Transition amplitudes}
For a multipole operator  
\begin{eqnarray}
\label{Mlaph}
{\cal M}(\lambda \mu) = \frac{1}{[\lambda]^{1/2}}
\sum_{rs} \langle r \parallel  {\cal M}_\lambda \parallel   s \rangle  \Bigl(a_r^\dagger \times b_s \Bigr)^{\lambda}_{\mu}, 
\end{eqnarray}
 the transition amplitudes are given by
\begin{equation}
\label{Mel}
\langle \psi_{\nu'} \parallel {\cal M} (\lambda) \parallel \psi_{\nu} \rangle 
=
 \sum_{n  n'} {\cal M}_{n n'  }^{\nu \nu'} (\lambda),  
\end{equation} 
where  
\begin{eqnarray}
\label{Mnnp}
{\cal M}_{n n'  }^{\nu \nu'} (\lambda) = \sum_{ \nu_n  \nu'_{n'}} {\cal C}^\nu_{\nu_n} {\cal C}^{\nu'}_{\nu'_{n'}}
\langle \nu'_{n'} \parallel {\cal M} (\lambda) \parallel  \nu_n \rangle 
\end{eqnarray}
for given  $n$ and  $n'$ and
\begin{eqnarray}
\label{Mnnpp}
\langle \nu'_{n'} \parallel {\cal M} (\lambda) \parallel  \nu_n \rangle =   [v]^{1/2}\nonumber\\   
\sum_{h \alpha_n h' \alpha'_{n'}} C^{\nu_n}_{h \alpha_n} 
{\cal M}^{\nu_n \nu'_{n'}}_{h \alpha_n h' \alpha'_{n'}} (\lambda) X^{\nu'_{n'}}_{h' \alpha'_{n'}}.
\end{eqnarray}
The matrix elements between the components with the same number  of phonons ($n' =n$) are given by 
 \begin{eqnarray}
\label{Mnnppp}
{\cal M}^{\nu_n \nu'_n}_{h \alpha_n h' \alpha'_{n}} (\lambda)=  \nonumber\\
(-)^{v' - v - \lambda} \delta_{\alpha_n \alpha'_n} 
  W(\lambda h v' \alpha_n; h' v) \langle h \parallel {\cal M}_\lambda \parallel h' \rangle 
\nonumber\\
- \delta_{h h'}   W(\lambda v' \alpha_n h; v \alpha'_n)
\langle \alpha'_n \parallel {\cal M} (\lambda) \parallel \alpha_n \rangle.   
\end{eqnarray}
For transitions between $n$ and $n'=n +1$ components  we have 
\begin{eqnarray}
{\cal M}^{\nu_n \nu_{n'}}_{h \alpha_n h' \alpha'_{n'}} (\lambda)=  - \delta_{h h'}
   W(\lambda v' \alpha_n h; v \alpha'_n)  
\nonumber\\
\times \sum_{x} {\cal M} \bigl(0 \rightarrow (x \lambda)\bigr) 
\langle \alpha'_{n'}\parallel O^\dagger_{(x\lambda)}  \parallel \alpha_n \rangle,
  \end{eqnarray}
where
\begin{equation}
{\cal M} \bigl(0 \rightarrow (x \lambda) \bigr)  =  \frac{1}{[\lambda]^{1/2}} \sum_{ph} \langle p  \parallel {\cal M}_\lambda \parallel h \rangle   
 c^{(x \lambda)}_{ph }  
\end{equation}
is just proportional to the TDA transition amplitude.

\begin{figure}[ht]
\includegraphics[width=\columnwidth ]{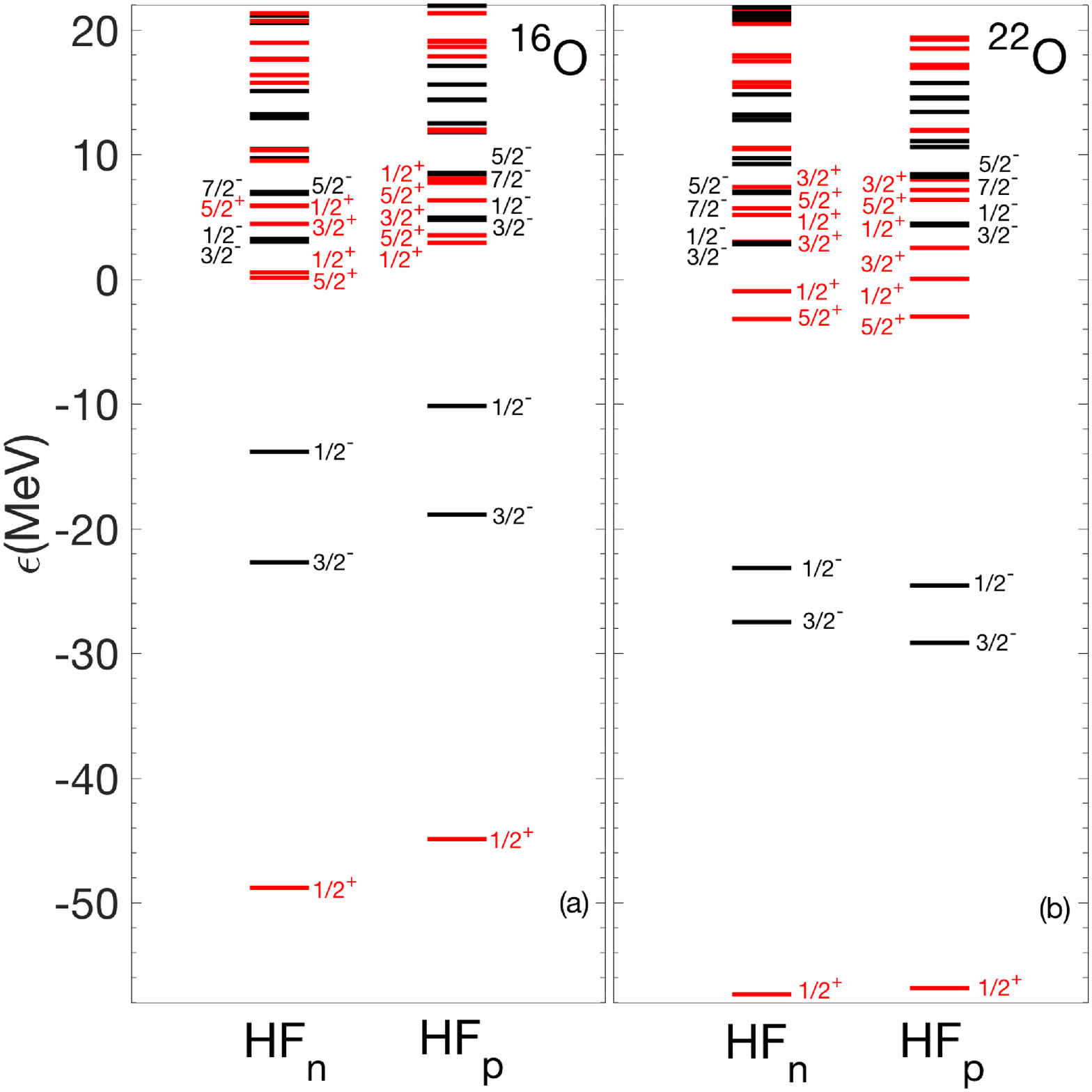}
\caption{(Color online) \label{fig1} HF spectra in $^{16}$O (a) and $^{22}$O (b)} 
\end{figure}

\begin{figure}[ht]
\includegraphics[width=\columnwidth ]{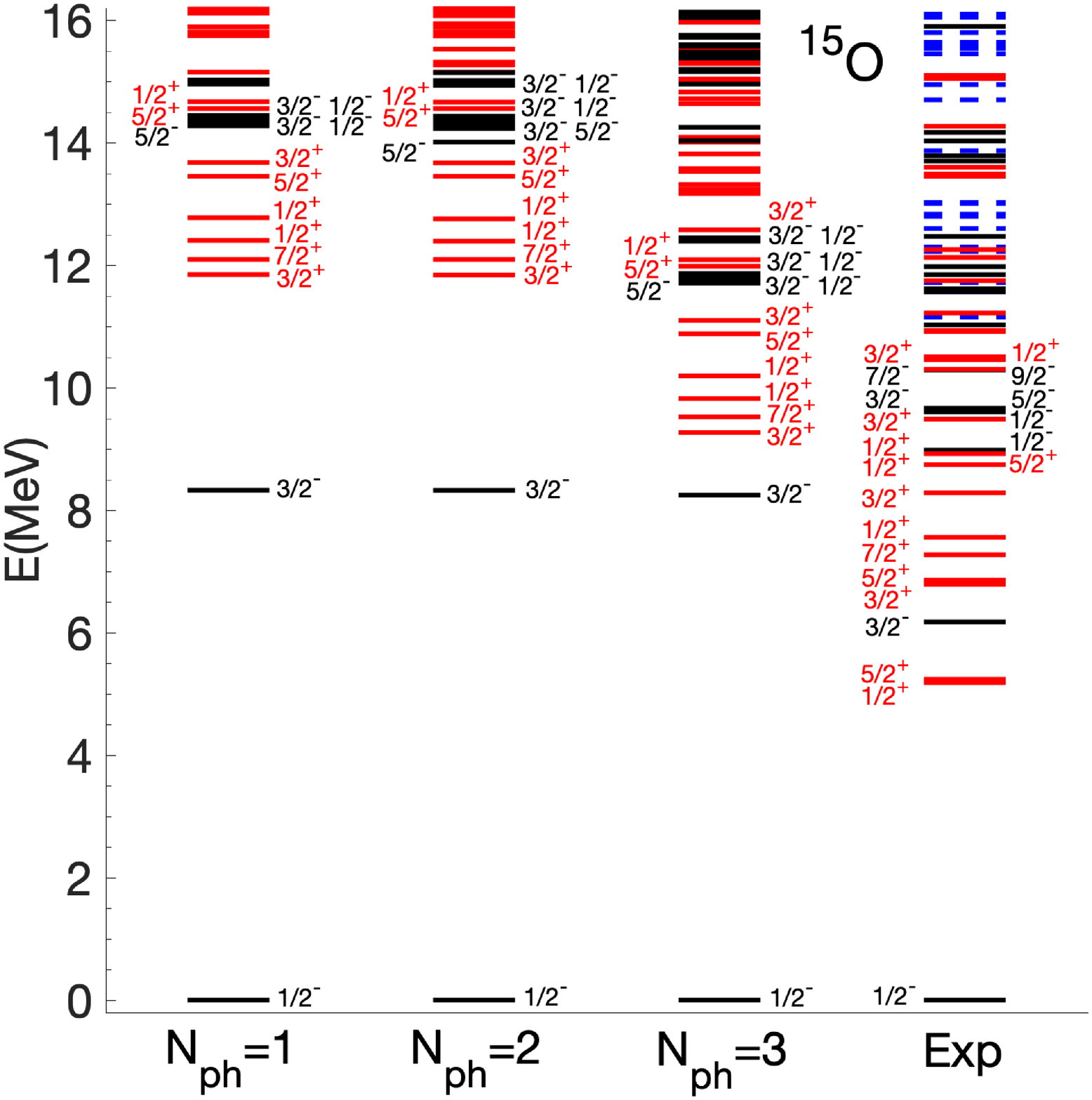}
\caption{(Color online) \label{fig2} Theoretical versus experimental \cite{AjzeSelo91} spectra of $^{15}$O.   The dashed levels have unknown spin or parity or both.} 
\end{figure}

\begin{figure}[ht]
\includegraphics[width=\columnwidth ]{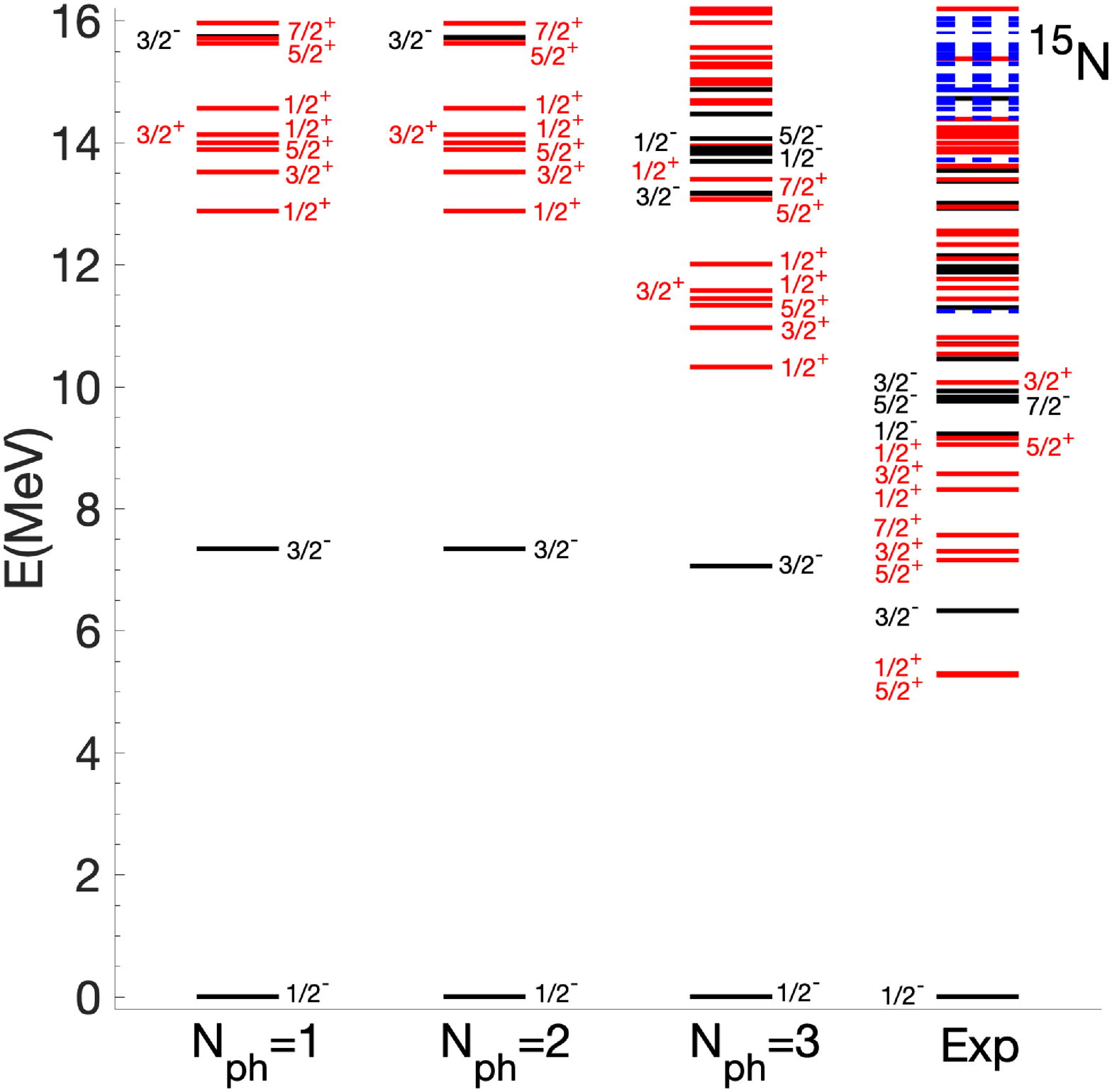}
\caption{(color online) \label{fig3} Theoretical versus experimental \cite{AjzeSelo91} spectra of  $^{15}$N.   The dashed levels have unknown spin or parity or both.} 
\end{figure}

\begin{table}  
\caption{\label{tab1} Phonon composition of selected states   (Eq. \ref{Psi}) of ${}^{15}$O and ${}^{15}$N.}
\begin{ruledtabular}
\begin{tabular}{ccccc}
\hline
&$\mid \nu \rangle$ & ${\cal E}_\nu$& $ (h^{-1} \times \lambda)$&$ W^\nu_{h \lambda}$ \\
\hline
$^{15}$O\\
\hline
&$ 1/2^-_1$&   0.000& $(1/2^- )$  &    96.63 \\
&$ 3/2^-_1$&   8.066& $(3/2^-)$  &    89.42\\
&$ 3/2^+_1$&  9.275&    $(1/2^- \times 1^-_1)$& 77.04\\
&$ 7/2^+_1$&   9.529&     $(1/2^- \times 3^-_1)$&89.16\\
&$ 1/2^+_1$&  9.830&     $(1/2^- \times 0^-_1)$&31.07\\
&                 &              &     $(1/2^- \times 1^-_1)$&55.05\\
&$ 5/2^+_1$&  10.885&      $(1/2^- \times 3^-_1)$&64.62\\
&                    &           &     $(1/2^- \times 2^-_1)$&10.61\\

&$ 3/2^+_2$&  11.105&     $(1/2^- \times 2^-_1)$&7.00\\
&                 &           &        $(3/2^- \times 3^-_1)$& 47.11\\
&                 &           &       $(3/2^- \times 3^-_2)$& 14.03\\
&$ 5/2^-_1$&   11.710&      $(1/2^- \times 2^+_1)$&85.51\\
&$ 3/2^-_2$&  13.104&     $(1/2^- \times 1^+_1)$&53.85\\
&                 &            &      $(1/2^- \times 2^+_1)$&32.63\\
&$ 5/2^-_3$&  15.474&      $(1/2^- \times 2^+_3)$&29.30\\
&                 &            &      $(1/2^- \times 3^+_2)$&27.40\\
&                 &            &      $(1/2^- \times 3^+_4)$&11.25\\
\hline
$^{15}$N\\
\hline
&$ 1/2^-_1$&   0.000& $(1/2^- )$  &   95.22\\
&$ 3/2^-_1$&   7.057& $(3/2^- )$  &   78.81 \\
&$ 1/2^+_1$&  10.321&$(3/2^- \times 2^-_1)$&6.35\\
&          &           &$(1/2^- \times 1^-_1)$&75.18\\
&          &           &$(1/2^- \times 1^-_2)$&5.46\\
&$ 3/2^+_1$&  10.963&$(1/2^- \times 1^-_2)$&70.60\\
& &   &                           $(3/2^- \times 3^-_1)$&11.49\\
&$ 5/2^+_1$&  11.331&  $(1/2^- \times 3^-_1)$&84.06\\
&$ 1/2^+_2$&   11.439&$(1/2^- \times 0^-_2)$&63.18\\
&          &           &$(1/2^- \times 1^-_2)$&18.39\\
&$ 7/2^+_1$&  13.153&   $(1/2^- \times 3^-_2)$&83.56\\
&$ 3/2^-_2$&  13.170&$(3/2^- \times 2^+_1)$&88.48\\
&$ 5/2^-_1$&  13.692&$(1/2^- \times 2^+_2)$&85.41\\
&$ 3/2^-_4$&  15.37&$(1/2^- \times 1^+_3)$&44.34\\
&          &           &$(1/2^- \times 2^+_2)$&37.65\\
 \end{tabular}
\end{ruledtabular}
\end{table} 

\begin{table}  
\caption{\label{tab2} Ground state magnetic moment $\mu$ ($\mu_N$)  and $\beta$-decay $ft$ value, $B(M1; J^\pi_i\rightarrow J^\pi_f$) (W.u.), and $B(E\lambda; J^\pi_i\rightarrow J^\pi_f$) (W.u.).
The experimental data are taken from Ref. \cite{AjzeSelo91}. }
\begin{ruledtabular}
\begin{tabular}{ccccc}
&& Th & Exp\\
\hline
$^{15}$O &$\mu$& +0.5986&$ \pm  0.7189 (8)$\\
&log$ft$&3.650&3.637\\ 
\hline
&$B(M1; 3/2^-_1 \rightarrow 1/2^-_1)$ &  0.56&$ >5.3 \times 10^{-2} $ \\
&$B(M1; 3/2^-_2 \rightarrow 1/2^-_1)$ & 0.001&$ 0.21$ \\

&$B(E2; 3/2^-_1 \rightarrow 1/2^-_1)$ & 0.03 &$ >0.28$ \\
&$B(E2; 5/2^-_1 \rightarrow 1/2^-_1)$ &  0.35&$ 15$ \\
&$B(E2; 5/2^-_3 \rightarrow 1/2^-_1)$ & 0.03 &$0.8\pm 0.5$ \\

&$B(E1; 1/2^+_1 \rightarrow 1/2^-_1)$& 0.06 &$(1.4 \pm 0.2) \times 10^{-3}$ \\
&$B(E1; 3/2^+_1 \rightarrow 1/2^-_1)$&   0.01&$>1.9 \times 10^{-4}$\\
&$B(E1; 3/2^+_2 \rightarrow 1/2^-_1)$&  0.05&$2.3 \times 10^{-3}$\\

&$B(E3;5/2^+_1 \rightarrow 1/2^-)$ & 3.95 &$4\pm 2 $ \\
&$B(E3; 7/2^+_1 \rightarrow 1/2^-_1)$ & 7.37&$6.4\pm 2.5$ \\
\hline  
$^{15}$N& $\mu$ & -0.249919471 & -0.283188842 (45) \\
\hline
&$B(M1; 3/2^-_1 \rightarrow 1/2^-_1)$ &  0.687 &$0.578 \pm 0.015$\\
&$B(M1; 3/2^-_2 \rightarrow 1/2^-_1)$ &  0.003&$(2.9 \pm 0.8)\times 10^{-2}$ \\
&$B(E2; 3/2^-_1 \rightarrow 1/2^-_1)$ &  1.20&$2.91  \pm 0.24$ \\
&$B(E2; 5/2^-_1 \rightarrow 1/2^-_1)$ & 0.37 &$1.3  \pm 0.3$ \\
&$B(E2; 3/2^-_4 \rightarrow 1/2^-_1)$ & 0.02 &$(2.4  \pm 0.6)  \times 10^{-2}$ \\
&$B(E1; 1/2^+_1 \rightarrow 1/2^-_1)$& 0.03&$(4.3  \pm 1.1)  \times 10^{-4}$\\
&$B(E1; 3/2^+_1 \rightarrow 1/2^-_1)$&  0.06 &$(6.7  \pm 0.05)  \times 10^{-2}$\\
&$B(E1; 1/2^+_2 \rightarrow 1/2^-_1)$&  0.01&$(1.3  \pm 0.8)  \times 10^{-3}$\\
&$B(E3;5/2^+_1 \rightarrow 1/2^-_1)$ &  3.11&$7  \pm 2$\\
&$B(E3; 7/2^+_1 \rightarrow 1/2^-_1)$ & 0.06 &$2.50  \pm 0.22$ \\ 
\hline
$^{16}$O&&&\\
&$B(E2;2^+_1\rightarrow0^+_1)$&0.379&3.1 $\pm$ 0.1\\
&$B(E2;2^+_2\rightarrow0^+_1)$&0.001&0.031 $\pm$ 0.003\\
&$B(E1;1^-_1\rightarrow0^+_1)$&0.014&$(3.5 \pm 0.2)\times 10^{-3}$\\
&$B(E1;1^-_2\rightarrow0^+_1)$&0.064&$(6.6 \pm 1.1)\times 10^{-5}$\\
&$B(E3;3^-_1\rightarrow0^+_1)$&3.19& 13.5   $\pm$ 0.7\\
&$B(E3;3^-_2\rightarrow0^+_1)$&2.21  &--- \\
 \end{tabular}
\end{ruledtabular}
\end{table} 
  
\section{Calculation details}
A Hamiltonian composed of   an  intrinsic kinetic operator $T_{int}$ and  the $NN$ optimized chiral potential $V_{NN}$ = NNLO$_{opt}$    \cite{Ekstr13}  was employed to generate the HF basis in a space encompassing  all harmonic oscillator shells up to $N_{max}=15$. 
 
The HF spectra for $^{16}$O and $^{22}$O are shown in Fig. \ref{fig1}.
One may notice the repulsive action of the Coulomb interaction on the proton single particle spectrum in $^{16}$O, which in $^{22}$O is counteracted by the strongly attractive interaction between the neutrons in excess and the protons.

The TDA basis was obtained using a subset of the HF states, spanning a space  encompassing up to $N=12$ for A=15 and $N=7$ for A=21.  We checked that the inclusion of higher energy shells does not affect the results.

The  $J^\pi = 1^-$ TDA phonons are free of  spurious admixtures induced by the center of mass (CM) motion. These spurious components have been removed by a method  discussed in Ref. \cite{Bianco14} based on the Gramm-Schmidt orthogonalization of the p-h  basis  to the CM state.

We used all one-phonon hole-core states $\mid (h^{-1} \times \alpha_1)^\nu \rangle$ in both A=15 and A=21 nuclei. 

In $^{15}$O and $^{15}$N, we included all the   states  $\mid (h^{-1} \times \alpha_2)^\nu \rangle$ of two-phonon energies $E_{\alpha_2} \leq 40$ MeV.

In $^{21}$O and $^{21}$N, we selected all  TDA phonons  having dominant $0-\hbar \omega$ and  $1-\hbar \omega$ components to build the two-phonon basis.

 \begin{figure}[ht]
\includegraphics[width=\columnwidth]{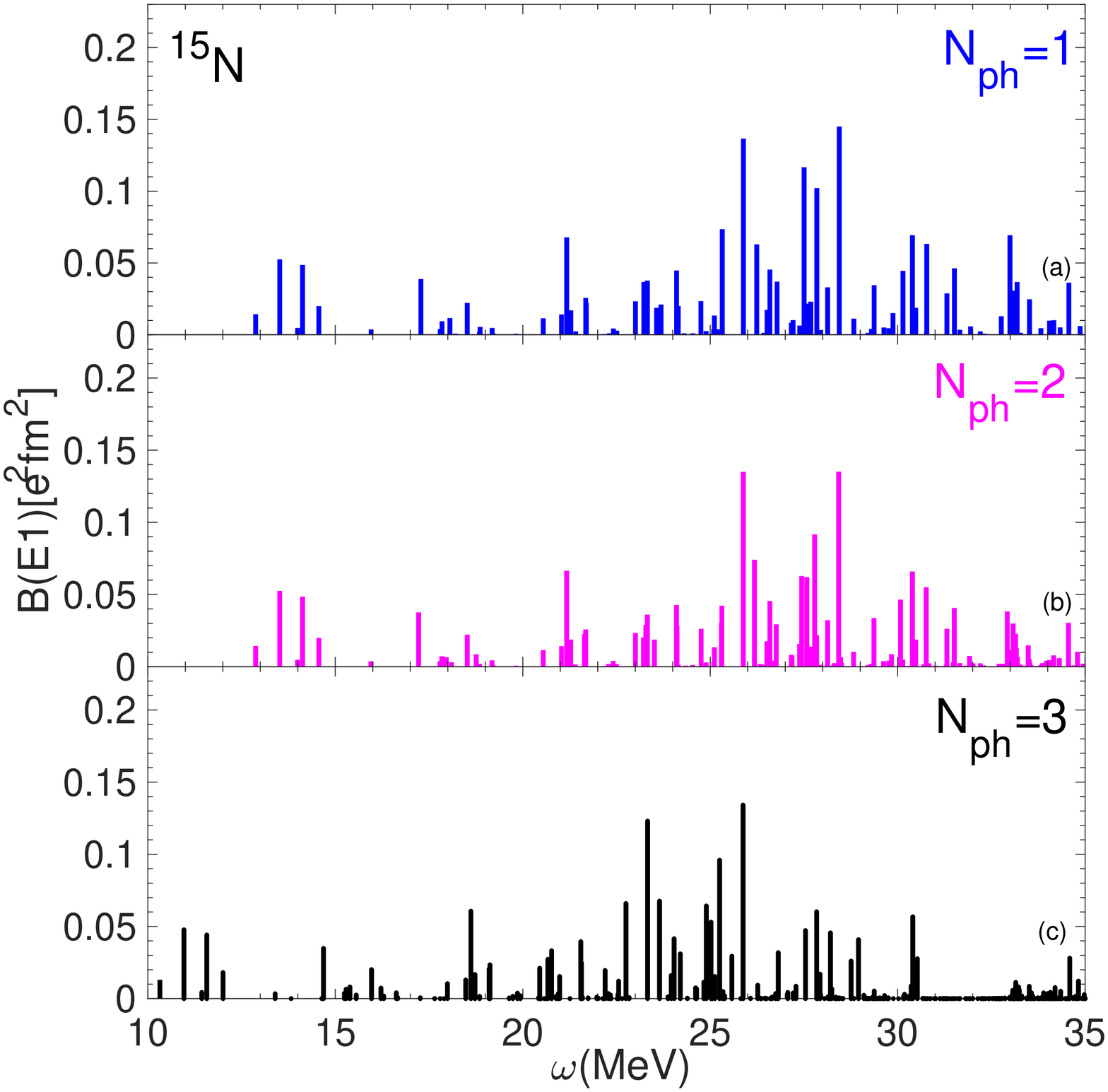}
\caption{(Color online) \label{fig4} $E1$ reduced strength distributions in different multiphonon spaces in  $^{15}$N.   
 }
\end{figure}

\begin{figure}[ht]
\includegraphics[width=\columnwidth ]{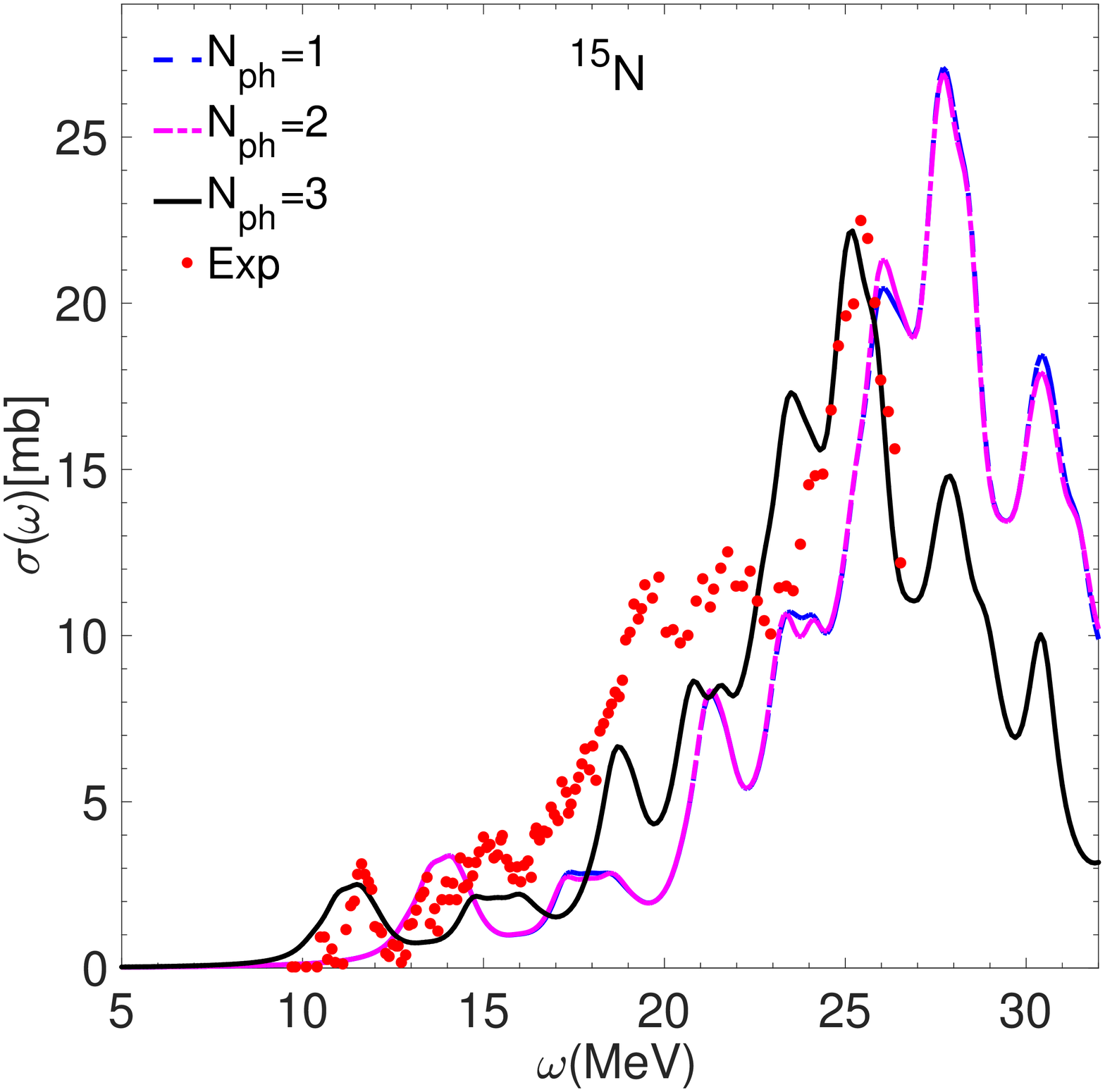}
\caption{(Color online) \label{fig5} The theoretical $E1$ cross sections, computed in different multiphonon spaces, are compared with the experimental ones \cite{Bates89} in  $^{15}$N. 
A Lorentzian of width $\Delta = 1$ MeV is used.} 
\end{figure}

In including the three-phonons we have neglected the interaction  between the one-phonon hole-core states $\mid \nu_1 \rangle$ and the two-phonons $\mid \alpha_2 \rangle$ so that $E_{\nu_3}  \sim E_{\nu_1} + E_\alpha$. We have included all phonons fulfilling the condition 
$E_{\nu_1} + E_\alpha \leq 65$ MeV. Furthermore, we have neglected the exchange terms between them in computing the  one-phonon to three-phonon coupling (Eq. (\ref{Vnn2})). Although this approximation may overestimate the coupling, the calculation should give a reliable indication of its importance.

The  reduced transition strengths $B(\lambda; \nu\rightarrow \nu')$ were computed using for the transition amplitudes the 
formula (\ref{Mel}) truncated up to $n=1$
\begin{eqnarray}
\label{Mel1}
\langle \psi_{\nu'} \parallel {\cal M} (\lambda) \parallel \psi_{\nu} \rangle 
=
 {\cal M}_{00  }^{\nu \nu'} (\lambda) + {\cal M}_{01 }^{\nu \nu'} (\lambda) + 
\nonumber\\
{\cal M}_{1 0  }^{\nu \nu'} (\lambda) + {\cal M}_{11}^{\nu \nu'} (\lambda).
\end{eqnarray}
The hole-hole piece is simply
 \begin{equation}
{\cal M}_{00  }^{\nu \nu'} (\lambda)  = (-)^{v + v' -\lambda} \sum_{h h'} {\cal C}^{\nu}_h {\cal C}^{\nu'}_{h'} \langle h \parallel {\cal M}_\lambda \parallel h' \rangle. 
\end{equation}
The hole-phonon transitions assumes also the simple form
 \begin{equation}
{\cal M}_{01 }^{\nu \nu'} (\lambda) =  - \sum_{h \nu'_1 x} {\cal C}^{\nu}_h  {\cal C}^{\nu'}_{\nu_1'} 
{\cal M} \bigl(0 \rightarrow (x \lambda) \bigr)  X^{\nu_1'}_{h (x \lambda)}.   
\end{equation}
${\cal M}_{10 }^{\nu \nu'} (\lambda)$ is easily deduced from the above formula. 

The phonon-phonon transition amplitudes ${\cal M}_{11}^{\nu \nu'} (\lambda)$  
are given by the general formulas (\ref{Mnnp} - \ref{Mnnppp}) for $n=n'=1$.

 We used the $E\lambda$ multipole operators 
\begin{equation}
\label{MEla}
{\cal M}(E\lambda \mu) = \sum_i e_i r_i^\lambda Y_{\lambda \mu} (\hat{r_i})
\end{equation}
 with bare charges $e_i =e$ for protons and $e_i =0$ for neutrons.

We have computed moments and  transition   strengths as well as the dipole cross section 
\begin{eqnarray}
\sigma (E1) &=&  \int_{0}^{\infty } \sigma (E1, \omega)   d\omega 
\nonumber\\
&=&\frac{16 \pi^3}{9\hbar c} \int_{0}^{\infty } \omega {\cal S}(E1, \omega) d\omega, 
 \label{sigma}
\end{eqnarray}
where ${\cal S}(E1, \omega)$ is the   strength function   
\begin{equation}
{\cal S}(E1, \omega) = \sum_{\nu } B_{\nu}  ( E1)\,\delta (\omega -\omega_{\nu })
\label{Strength}
\end{equation}
and  $B_\nu (E1) = B (E1 ; g.s. \rightarrow \nu)$ is the reduced strength of the transition to the $\nu_{th}$ final state of energy  $\omega_\nu = {\cal E}_\nu - {\cal E}_{g.s.}$.  
In practical calculations  the $\delta$ function is replaced by a Lorentzian  of width $\Delta$.

For the magnetic dipole moment and the $M1$ transitions we adopted the operator
\begin{equation}
\vec{\mu} = \sum_k \Bigl(g_l (k) \vec{l}_k  + g_s (k) \vec{s}_k \Bigr)
\end{equation}
with  bare gyromagnetic factors, $g_l(k)=1$ and $g_s (k) = 5.59$ for protons, $g_l (k) =0$ and $g_s (k) = - 3.83$ for neutrons.

For the  $\beta$-decay transitions  we used the Fermi and Gamow-Teller operators
\begin{eqnarray}
{\cal M}_F = g_V \sum_k t_\pm (k), \\
{\cal M}_{GT} = g_A \sum_k t_\pm (k) \vec{\sigma}(k)
\end{eqnarray}
with  the bare weak charges $g_V =1$ and  $g_A =1.25$. We have introduced the spherical components $t_\mu$ of the isospin single particle operator.

Different formulas hold  for  $\beta$-decay transition amplitudes.
The hole-hole components are given by 
\begin{equation}
\label{M00b}
{\cal M}_{00 }^{\nu \nu'} (\lambda) = (-)^{v  - v' -\lambda} \sum_{i j} {\cal C}^{\nu}_{v_i} {\cal C}^{\nu'}_{v_j'}   
  \langle v_i \parallel {\cal M}_\lambda \parallel v_j' \rangle. 
\end{equation}
For the hole-phonon pieces we have
\begin{equation}
\label{M01b}
{\cal M}_{01 }^{\nu \nu'} (\lambda)= 
\sum_{i \nu'_1} C^{\nu}_{v_i}  C^{\nu'}_{\nu_1'}
\langle \nu_1'  \parallel {\cal M} (\lambda) \parallel v_i^{-1} \rangle,
\end{equation}
where
\begin{equation}
\label{M01b1}
\langle \nu_1'  \parallel {\cal M} (\lambda) \parallel v_i^{-1} \rangle
= \sum_{ph'  } \langle p \parallel {\cal M}_\lambda  \parallel h' \rangle W_{p h'}^{\nu \nu'}(\lambda) 
\end{equation}
and
\begin{eqnarray}
\label{M01b2}
W_{p h'}^{\nu \nu'}(\lambda) =  (-)^{v + p + v' + h'} \sum_\sigma [\sigma]^{1/2} c^{\sigma}_{p v_i}
\nonumber\\ 
\times W(v' h' v p; \sigma \lambda) \langle \sigma \parallel a^\dagger_{h'} \parallel \nu'_1 \rangle.           
\end{eqnarray}
The phonon-phonon terms ${\cal M}_{11}(\lambda)$ are given by
\begin{eqnarray}
\label{M11b}
{\cal M}_{11}(\lambda) =   \sum_{ \nu_1 \nu_1'} {\cal C}^{\nu}_{\nu_1} C^{\nu'}_{\nu_1'}  
\langle \nu_1'  \parallel {\cal M} (\lambda) \parallel \nu_1 \rangle,          
\end{eqnarray}
 where
\begin{equation}
\label{M11b1}
\langle \nu_1'  \parallel {\cal M} (\lambda) \parallel \nu_1 \rangle =
\sum_{h_\pi h'_\nu} \langle h_\pi \parallel {\cal M}_\lambda \parallel h'_\nu \rangle W_{h h'}^{\nu \nu'}(\lambda)
\end{equation}
and
\begin{eqnarray} 
\label{M11b2}
  W_{h h'}^{\nu \nu'}(\lambda) = 
 \sum_{\sigma } (-)^{v'_\nu + h'_\nu - \sigma} 
\nonumber\\
W(v'h'_\nu v h_\pi; \sigma \lambda) 
\langle \sigma \parallel a^\dagger_{h'_\nu} \parallel \nu'_1 \rangle
\langle \sigma \parallel a^\dagger_{h_\pi} \parallel \nu_1 \rangle.       
\end{eqnarray}

\section{spectroscopy of  $^{15}$O and $^{15}$N}

\subsection{Spectra and phonon composition of the wavefunctions }
\label{spectraON15}
The theoretical spectra of $^{15}$O  and  $^{15}$N computed in different multiphonon spaces  are compared   with the experiments  
in Figs. \ref{fig2} and \ref{fig3}, respectively.

The lowest level is the $3/2^-_1$, the only single-hole excited state (Tab. \ref{tab1}). It is $\sim 0.7$ MeV above the corresponding experimental level  in $^{15}$N and  $\sim 2$ MeV above in $^{15}$O.

 All the other states arise from the excitations of the core and are too high in energy.
The lowest levels have spin compatible with the experimental ones but are   $\sim 4$ MeV and $\sim 5$ MeV above in $^{15}$O and $^{15}$N, respectively.

These large gaps originate from the too  high     TDA phonon energies which weaken  the phonon coupling and, therefore,
induce small admixing among different phonon components. 

In fact, the   wavefunctions of the low lying levels have a dominant one-phonon character with $\sim 10 \%$ admixing of three phonons 
(Table \ref{tab1}).    
 
It is even more important to notice the strong violation of the mirror symmetry between $^{15}$O and $^{15}$N in  disagreement with the experiments and theoretical analyses based on shell model \cite{Alburger79,Raman94}. The two spectra, in fact, differ in the energies and sequence of  few low-lying levels and in the structure of the corresponding wavefunctions (Tab. \ref{tab1}). The asymmetry is especially pronounced in states like the $3/2^+_1$ and $7/2^+_1$, where the Pauli principle plays a crucial role.

This anomaly is due to the combined effect of  the charge symmetry breaking of the two-body potential and the enforcement of the Pauli principle of the hole-phonon basis through the Cholesky decomposition method.

In order to illustrate the problem we consider the lowest $7/2^+_1$ of  $^{15}$O and $^{15}$N. 
In shell model, the dominant lowest energy  configuration entering $7/2^+_1$ in $^{15}$O is 
$\bigl[(0p^{-1}_{1/2} (\nu) \times 0p^{-1}_{1/2} (\pi))^1 \times 0d_{5/2} (\pi) \bigr]^{7/2}$. 
The configuration $\bigl[(0p^{-1}_{1/2} (\nu) \times 0p^{-1}_{1/2} (\nu))^1 \times 0d_{5/2} (\nu) \bigr]^{7/2}$ is excluded by the Pauli principle. A similar argument holds for $^{15}$N. We need just to interchange neutrons and protons.

In our hole-phonon scheme, the $7/2^+_1$ arises from coupling the neutron or proton hole
$0p^{-1}_{1/2}$ to the low-lying $3^-$ phonons.  
The two lowest     $3^-_1$ and $3^-_2$ are $\sim 3.5$ MeV far apart.  This large splitting  is caused by the differences ($\sim 1$ MeV) between the HF proton and neutron energy separations (Fig. \ref{fig1}), produced by the charge symmetry breaking terms of the two-body potential, and amplified by the strong proton-neutron interaction.
The two $3^-$ states are linear combinations of   proton   and neutron  p-h configurations and  have proton and neutron dominance, respectively. The protons (neutrons) account for  $\sim 60 \%$ ( $\sim 40 \%$) of the $3^-_1$ and $\sim 40 \%$ ( $\sim 60 \%$) of the $3^-_2$.

The Cholesky decomposition method selects the hole-phonon components rather than the single p-h terms as in shell model. In the specific example,  it selects  the   $[0p^{-1}_{1/2}(\nu) \times 3^-_1]^{7/2}$ state for $^{15}$O (Table \ref{tab1}) since the $3^-_1$ has a proton dominance and discards as redundant the $[0p^{-1}_{1/2}(\nu) \times 3^-_2]^{7/2}$ which contains the neutron dominant $3^-_2$ phonon.  On the contrary, it selects  the   $[0p^{-1}_{1/2}(\pi) \times 3^-_2]^{7/2}$ state for  $^{15}$N and discards as redundant the $[0p^{-1}_{1/2}(\pi) \times 3^-_1]^{7/2}$.
The $7/2^+_1$ states so selected, however, are   $\sim 3.5$ MeV far apart, because of the energy splitting between the two $3^-_1$ and $3^-_2$ phonons, and have different structure because of the different proton-neutron content. The mirror symmetry is broken thereby.

Such a symmetry is preserved only if we  turn off the Coulomb potential and neglect the mass differences between protons and neutrons. This charge symmetric interaction yields identical proton and neutron HF spectra and TDA states with equal proton and neutron content ($50 \%$). In this case Cholesky selects the   same   hole-phonon basis states $\mid (h^{-1} \times \lambda)^v \rangle$  for both nuclei and, therefore, yields identical spectra and wavefunctions. 

On the other hand, even a small deviation from $50 \%$ of the proton-neutron content breaks the mirror symmetry. For instance, if the proton content of the $3^-_1$ ($3^-_2$) is slightly larger (smaller) than  $50 \%$,    the Cholesky method selects the state   $[0p^{-1}_{1/2}(\nu) \times 3^-_1]^{7/2}$ for $^{15}$O and  $[0p^{-1}_{1/2}(\pi) \times 3^-_2]^{7/2}$ for $^{15}$N (Table \ref{tab1}). These two states, however, remain  far apart in energy, because of the energy splitting between the two $3^-_1$ and $3^-_2$ induced by the proton-neutron interaction, with consequent breaking of the mirror symmetry.   

Since it is very unlikely that an exactly equal proton-neutron content can be obtained for any potential, we need to modify our hole-phonon scheme. We may, for instance, neglect the charge symmetry breaking terms in generating the HF and TDA basis states and reintroduce them directly in the eigenvalue equations (\ref{eigChnew}) for odd nuclei.   
   
\subsection{Moments and transitions}
The magnetic moment of $^{15}$O  is fairly close to the corresponding experimental value (Table \ref{tab2}). The agreement is even better 
for $^{15}$N.
The {\it ft} value of the ground state $\beta$-decay of $^{15}$O is well reproduced. It gets contribution from both Fermi and Gamow-Teller
transitions with respective strengths $B_F = 0.895$ and $B_{GT}= 0.480$.

The   $M1$ $3/2^-_1 \rightarrow 1/2^-_1$ decay transition  is basically of single hole nature. Its reduced strength  is    fairly close to the measured value in $^{15}$N and is compatible with the lower limit established experimentally for $^{15}$O.
 The    $M1$ $3/2^-_2 \rightarrow 1/2^-_1$  transitions involve a one-phonon hole-core initial state in both nuclei. They  are determined by the $1^+_1$ and  $1^+_3$ phonons in $^{15}$O and $^{15}$N, respectively.   The experimental strengths  are  one order and two orders of magnitude larger than the corresponding theoretical values     in $^{15}$N and  and $^{15}$O, respectively.

The  $E2$ $3/2^-_1 \rightarrow 1/2^-_1$ transition is too weak in $^{15}$O since the dominant single-hole neutron components do not contribute.  In $^{15}$N, instead, it is much stronger because of the contribution coming from the proton single-hole components. Its strength is   less than half  the experimental value.

The  $E2$ $5/2^-_1 \rightarrow 1/2^-_1$ transitions involve a one-phonon hole-core state and are basically determined by the 
$2^+_1$ phonon in $^{15}$O and   the $2^+_2$ in $^{15}$N. The strengths of the $E2$ transition from these two TDA phonons to the HF ground state
are orders of magnitude smaller than the corresponding experimental values in $^{16}$O (Table \ref{tab2}). 
The numbers in the Table explain why the $B(E2)$ is larger in  $^{15}$O than in $^{15}$N and why the calculation underestimates the experimental values in both nuclei by orders of magnitude.

 The $E1$  strengths are larger than the experimental values in $^{15}$O. The transitions are entirely determined by the $1^-$ phonons. The amplitudes of the  hole-core $(h^{-1} \times 1^-_1)^\nu$ components of the low-lying $3/2^+_1$ and $1/2^+_1$ states   have got enhanced by the coupling to the tree-phonon hole-core states and have become   dominant   in those states (Table \ref{tab1}).
The $E1$ transition strengths of 	$^{15}$N are comparable with those computed for  $^{15}$O and with the measured data.  	
	 
The $E3$ transition strengths are comparable with the experimental quantities in both nuclei except for the $B(E3; 7/2^+_1 \rightarrow 1/2^-_1)$ in $^{15}$N which is orders of magnitude smaller that the measured value. This weak decay can be only partially explained by the fact that the $3^-_2$ phonon, which determines this transition,  is more weakly coupled to the $0^+_1$ ground state than the $3^-_1$ phonon which induces the other transitions.   
 
The role of the multiphonon states emerges clearly from the analysis of the dipole response.
As shown in  Fig. \ref{fig4} for $^{15}$N, both two-phonon and three-phonon components exert a weak quenching action. The coupling to three phonons, however, induces a  shift of the strength toward the lower sector of the spectrum thereby yielding a fair agreement between theoretical and experimental \cite{Bates89} cross sections  (Fig. \ref{fig5}).

The cross section, integrated up to  up 40 MeV,  exhausts $\sim$112\% of the Thomas-Reike-Khun (TRK) sum rule. The sum up to 26.5 MeV exhausts 50\% of the TRK sum which is fairly close to the value  (58\%) deduced from the experimental data.

\begin{figure}[ht]
\includegraphics[width=\columnwidth]{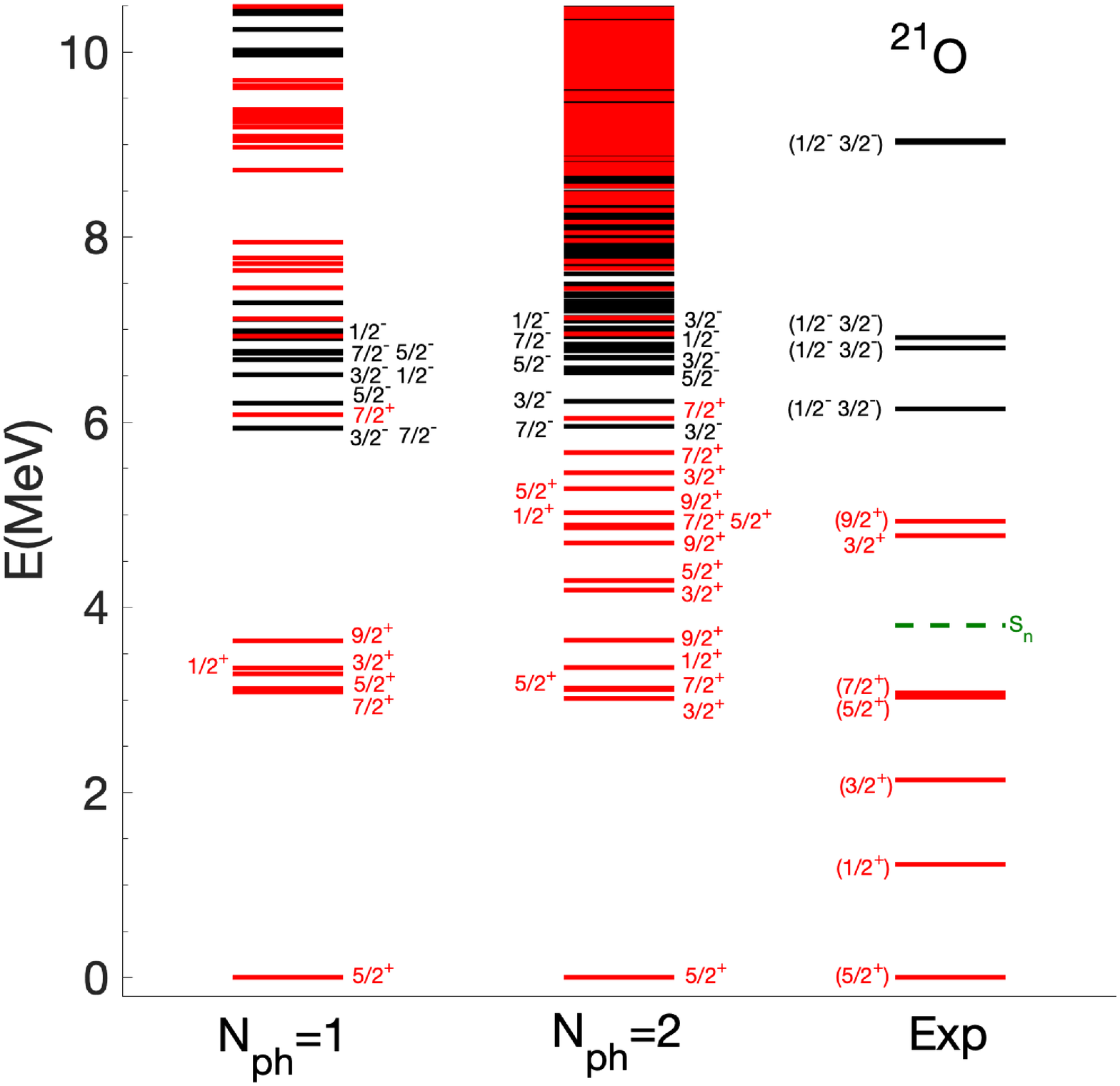}
\caption{(Color online) \label{fig6}Theoretical versus experimental \cite{Firestone15} spectra of $^{21}$O.    }
\end{figure}
\begin{figure}[ht]
\includegraphics[width=\columnwidth ]{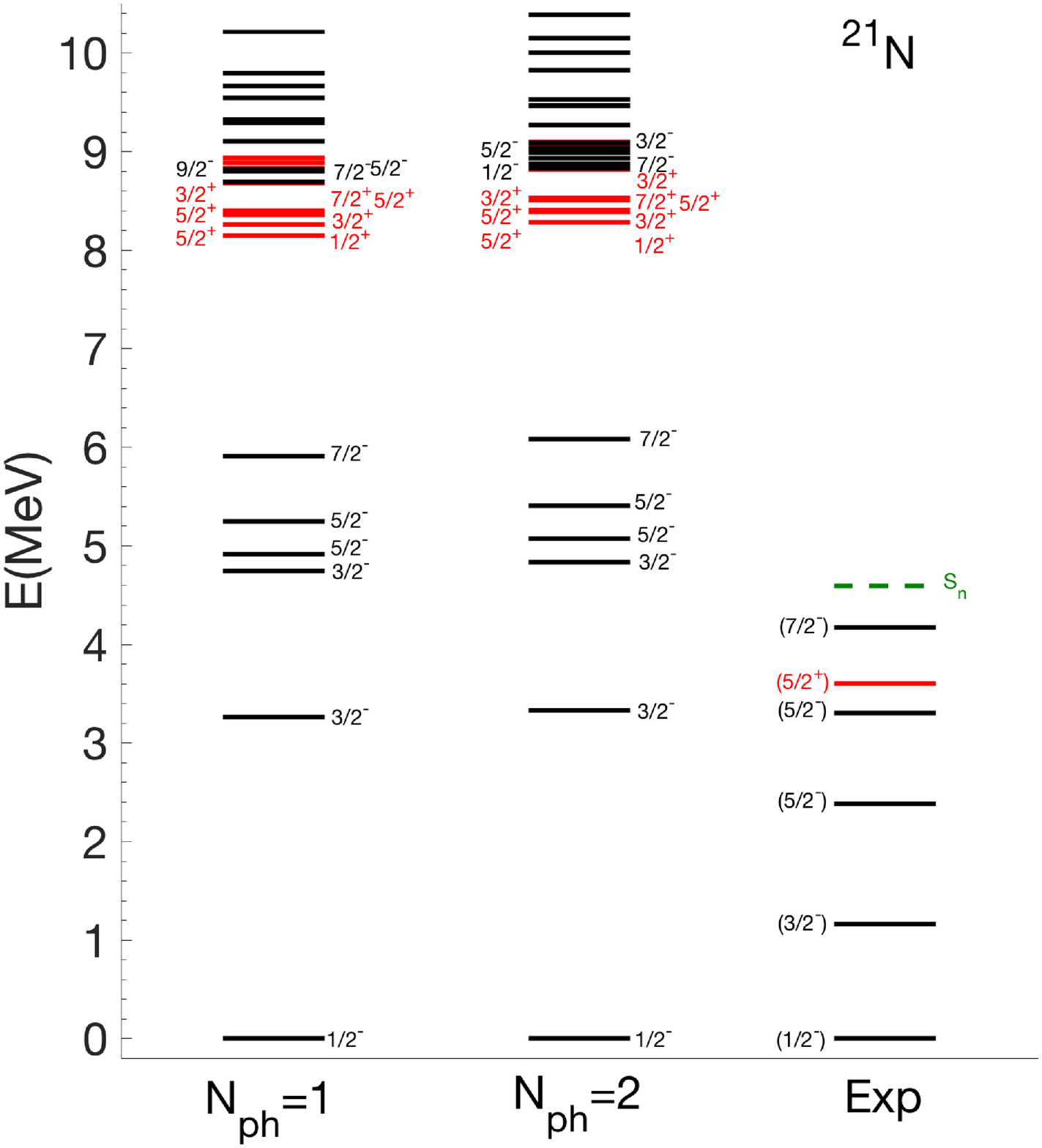}
\caption{(Color online) \label{fig7} Theoretical versus experimental \cite{Firestone15} spectra of  $^{21}$N. } 
\end{figure}

 \begin{table}  
\caption{\label{tab3}Phonon composition of selected states   (Eq. \ref{Psi}) of ${}^{21}$O and ${}^{21}$N. }
\begin{ruledtabular}
\begin{tabular}{ccccc}
\hline
&$\mid \nu \rangle$ & ${\cal E}_\nu$& $ (h^{-1} \times \lambda)$&$ W^\nu_{h \lambda}$ \\
\hline
$^{21}$O\\
&$5/2^+_1$ &  0.000 & $(5/2^+  )$&97.02 \\
&$3/2^+_1$ &3.0115 &$(5/2^+ \times 2^+_1)$&75.00\\
&$7/2^+_1$ &3.1099 &$(5/2^+ \times 2^+_1)$&99.71\\
&$5/2^+_2$  &3.1251 &$(5/2^+ \times 3^+_1)$&98.00\\
&$1/2^+_1$ & 3.3481 &$(5/2^+ \times 3^+_1)$&99.50\\
&$9/2^+_1$  &3.6425 &$(5/2^+ \times 3^+_1)$&98.01\\
&$3/2^-_1$ &5.9520 &$(5/2^+ \times 3^-_1)$& 90.71\\
&           &      &$(5/2^+ \times 2^-_2)$&7.30\\

&$3/2^-_2$ &$6.5306$&$(5/2^+ \times 4^-_1)$&81.35\\
&          &       &$(5/2^+ \times 2^-_2)$&15.93\\ 
&$1/2^-_1$ &6.6922 &$(5/2^+ \times 2^-_1)$&72.57\\
&           &      &$(5/2^+ \times 2^-_2)$&26.92\\

&$1/2^-_4$ &7.3024 &$(5/2^+ \times 2^-_1)$&24.99\\
&           &      &$(5/2^+ \times 2^-_2)$&67.43\\
\hline
$^{21}$N\\
&$1/2^-_1$   & 0.000   &  $(1/2^-  )$&83.65 \\
&$3/2^-_1$   &  3.3267 &  $(3/2^-  )$& 72.58 \\
&$ 3/2^-_2$    &  4.8317  &$(1/2^- \times 2^+_1)$& 89.40\\
&$ 5/2^-_1$    & 5.0700 &$(1/2^- \times 2^+_1)$&95.20\\
&$ 5/2^-_2$    & 5.4029 &$(1/2^- \times 3^+_1)$&88.50\\
&          &       &$(3/2^- \times 3^+_1)$&6.41\\
&$ 7/2^-_1$ & 6.0796 &$(1/2^- \times 3^+_1)$&97.67\\
&$ 1/2^+_1$  & 8.2790 &$(1/2^- \times 1^-_1)$&92.46\\
&$ 3/2^+_1$  &8.3851  &$(1/2^-  \times 2^-_1)$ &75.86\\
&          &       &$(1/2^- \times 1^-_1)$&15.34\\

&$ 5/2^+_1$ &8.4037  &$(1/2^- \times 3^-_1)$&78.38\\
&          &       &$(1/2^- \times 2^-_1)$&9.56\\
&$ 7/2^+_2$& 8.5099  &$(1/2^- \times 4^-_1)$&78.45\\
&          &       &$(1/2^- \times 3^-_1)$&11.29\\
&          &       &$(3/2^- \times 4^-_1)$&5.40\\

 \end{tabular}
\end{ruledtabular}
\end{table}

\begin{table}  
\caption{\label{tab4}Ground state $\beta$-decay of $^{21}$N. The experimental data are taken from Ref. \cite{Li09}.
The spins  of the final states have not been determined experimentally.}
\begin{ruledtabular}
\begin{tabular}{cccccc}
 &$\nu_{f}$&$E^{f}$&log$ft$&B(GT)\\
\hline
EMPM\\
&$3/2^-$&5.95  &7.14 & 0.00044    \\
&$3/2^-$&6.53  &7.62 & 0.00015    \\
&$1/2^-$&6.69  &8.29   & 0.00003  \\
&$1/2^-$&7.30  &7.55 & 0.00016    \\
&$3/2^-$&10.02  &5.60 & 0.01513    \\
&$3/2^-$&10.43  &6.30 & 0.00306    \\
&$3/2^-$&13.05 &5.70 & 0.01221    \\
&$3/2^-$&14.54 &5.32 & 0.02910    \\
&$1/2^-$&14.70 &5.57 & 0.00108    \\
&$3/2^-$&17.77 &4.70 & 0.12168    \\
&$1/2^-$&18.84 & 5.77& 0.00987    \\
&$3/2^-$&20.86 &4.33 & 0.28512    \\
&$1/2^-$&22.45 &4.23 & 0.20025     \\
&$1/2^-$&22.84 &4.32 & 0.27221    \\
\hline
Exp\\
&$ (1/2^-, 3/2^-)  $  &6.14   &$5.44\pm 0.06$  & $0.0224\pm 0.0032$    \\
&$ (1/2^-, 3/2^-)  $  &6.80   &$5.19\pm 0.06$ & $0.0399\pm 0.0056$      \\
&$ (1/2^-, 3/2^-)  $  &6.91   &$5.44\pm 0.07$&  $0.0224\pm 0.0035$     \\
&$ (1/2^-, 3/2^-)  $  &9.02   &$4.78\pm 0.06$&$0.1015\pm 0.0145$        \\
&$ (1/2^-, 3/2^-)  $  &9.04   &$4.62\pm 0.06$&$0.1462\pm 0.0206  $      \\
 \end{tabular}
\end{ruledtabular}
\end{table}

\section{Spectroscopy of $^{21}$O and $^{21}$N}

As shown in Fig. \ref{fig6},   the theoretical spectrum of $^{21}$O is very dense,  covers large part of the experimental region, and contains several levels of spins compatible with those attributed to the  experimental ones. It misses, however, the two low-lying levels detected experimentally.

 As in $^{22}$O and $^{23}$O \cite{DeGreg18}, all low-lying states have substantially a single $n$-phonon structure (Table \ref{tab3}). They are determined by the excitation of the neutrons in excess which are governed by the weak neutron-neutron interaction. Moreover,    the  Pauli principle exerts an inhibiting action.

The $5/2^+_1$  ground state has  a single-hole nature.   
The next five levels are in the energy interval $\sim 3.0-4.0$ MeV and form a quintuplet of positive parity states  $\{1/2^+_2,3/2^+_2,5/2^+_2,7/2^+_1,9/2^+_1\}$   built by coupling the $5/2^+_1$  neutron hole to the low-lying one-phonon states  
$2^+_1$ and $3^+_1$ occurring in $^{22}$O (Table \ref{tab3}). 

A sequence of two-phonon hole-core states  up to $\sim 6$ MeV follow.
Thus, like $^{23}$O,  the $^{21}$O theoretical level scheme retains the harmonic nature of the spectrum predicted, but only partially confirmed experimentally, for $^{22}$O \cite{DeGreg18}. Only few members of those multiplets appear in the the experimental spectrum.   

The first negative parity states occur at $\sim 6$ MeV and are in correspondence with the experimental levels of the same parity. They have in general a dominant configuration in which a $5/2^+$ hole couples mostly to $2^-$ phonons and in 
few cases to $3^-$ and $4^-$ (Table \ref{tab3}).

Theoretical and experimental spectra of $^{21}$N are shown in  Fig. \ref{fig7}.
 The   ground and  first excited states have a dominant HF component admixed appreciably with the one-phonon hole-core states 
(Table \ref{tab3}). This admixing is due to the strong interaction between the proton hole and the neutrons in excess which 
induces a strong hole-phonon coupling   and shifts downward the energies with respect to the other low-lying states.    

These, in fact, have an almost pure one-phonon character since their coupling with the two-phonon components is governed  by the weak neutron-neutron interaction.  The $3/2_2^-$ and first $5/2^-_1$ are composed  of a $1/2^-$ hole coupled to the $2^+_1$ phonon, while the second $5/2^-_2$ and $7/2^-_1$  are built of a $1/2^-$ hole coupled to the $3^+_1$ phonon (Table \ref{tab3}), in agreement with the shell model analysis reported in \cite{Firestone15}.
These states are in one to one correspondence with the available experimental levels but fall at too high energies. In fact, they keep their unperturbed TDA energies and, moreover, get more distant from  the dominantly HF ground and first excited states which are shifted downward by the hole-phonon coupling.   
 
The $ft$ values of few  $\beta$-decay transitions   are the only additional experimental data available for  $^{21}$N  
\cite{Li09}. The states of $^{21}$O  populated by these decays are in the energy interval $\sim6 - 9$ MeV (Table \ref{tab4}). Their spins were not uniquely determined. Only the values $1/2^-$ or $3/2^-$ are compatible with the ground state spin of $^{21}$N. 

In our calculation, only states of too  high energy get populated with a rate comparable with the data. The low-lying states falling in the energy region of observation are poorly populated. In fact, the low-lying   $1/2^-$ or $3/2^-$ states of $^{21}$O  have a hole-phonon character and cannot be populated through the hole-hole transition ${\cal M}_{00}$ (\ref{M00b}). 

The hole-phonon transition amplitudes ${\cal M}_{01}$ (\ref{M01b}) are also small (see Eqs. (\ref{M01b})-(\ref{M01b2})). In fact,  the low-lying TDA phonons which are dominant in the mentioned   $1/2^-$ or $3/2^-$ are composed mainly of the neutron configurations 
 $((1p) \times (0d5/2)^{-1})_\nu^{\sigma}$. Therefore, the coefficients of the proton p-h components 
$((0d_{5/2}) \times (0p)^{-1})^\sigma_\pi$ contributing to the strength through the   weight (\ref{M01b2}) are very small and suppress the transition amplitudes. 

Similarly, the small coefficients of both proton and neutron p-h configurations $((0d_{5/2}) \times (0p)^{-1})^\sigma$ suppress the hole-hole
transition amplitudes ${\cal M}_{11}$ (see Eqs. (\ref{M11b})-(\ref{M11b2})).

\section{Concluding remarks}

Let us enumerate the most meaningful results of our comprehensive comparative analysis: i) In both $^{15}$O and $^{15}$N,  all low-lying  hole-core states have  energies several MeV above the corresponding  experimental levels and do not reproduce the mirror symmetry observed experimentally. They have a dominant one-phonon character and couple strongly only to three phonons. In virtue of such a coupling, the theoretical $E1$ cross section falls in the region of observation in $^{15}$N and has shape and magnitude in fair agreement with the experimental quantity.  ii) A less marked energy gap between excited and ground states occurs  in the neutron rich  $^{21}$N. In this nucleus, the low-lying states are in  one-to-one correspondence with the available experimental levels. In $^{21}$O, the theoretical spectrum   overlaps to a large extent with the experimental one but fails to reproduce the lowest two levels. iii) In all nuclei, the  low-lying states have an almost pure one-phonon hole-core nature.  The coupling to two-phonon  states is ineffective and the three-phonon components do not promote a sufficient downward shift of their energies.  

 The violation of the mirror symmetry is induced by the different selection of the basis states extracted by the Cholesky method from the redundant hole-phonon basis of $^{15}$O and
 $^{15}$N . This is a serious limitation of the hole-phonon scheme, which affects a few low-lying states in any case. It can be overcome by neglecting the charge symmetry violating terms in generating the HF and TDA basis and including them directly in the hole-phonon eigenvalue equation.  

The too high energy and the pure one-phonon nature of the low-lying states have a common origin. The  energy separation between the $(sd)$ and the $(0p)$ HF states is too large, as illustrated in Fig. \ref{fig1}. In $^{15}$O, $^{15}$N, and to a less extent $^{21}$N, this large gap yields TDA phonons  of too high energies and, consequently, large gaps between  different  $n$-phonon subspaces thereby weakening the coupling between them. 
The large $(sd)$-$(0p)$ energy separation inhibits the presence of protons in the low-lying TDA constituent phonons of the hole-core states in $^{21}$O.  These have therefore a neutron character and the coupling with the other $n$-phonon subspaces, being governed by  the weak neutron-neutron interaction, is very weak. Hence the $n$-phonon purity of the states.

On the other hand, one needs  a strong coupling in order to push down in energy the low-lying levels and to enhance the strength of some transitions, especially the electric quadrupole and the $\beta$-decay transitions.

The recipe is the same suggested by analogous investigations of the  A=17 \cite{DeGreg17a} and A=23 \cite{DeGreg18} nuclei of the same region with a valence particle:
Only smoother HF spectra in the low-energy sector yielding TDA phonons of lower energy can induce   a   more effective coupling between different $n$-phonon subspaces and, therefore, an appreciable phonon mixing  in the low-lying states.

This recipe holds even if we were able to include four or six phonons. In fact, even in phenomenological shell model calculations,    the low-lying positive parity levels in $^{16}$O could be described with a fair approximation in a space including up to 4p-4h configurations only after assuming a substantially reduced separation between the $(sd)$ and $0p$ shells \cite{HaxJoh90,WarBrowMil92}. An equally small gap was necessary  in order to describe the low-lying levels of odd nuclei around $^{16}$O in an analogous shell model calculation which included up to 6p-6h configurations \cite{Utsuno2011}.

A new interaction is needed in order to generate a smoother HF spectrum.   A somewhat compact  spectrum can be roughly obtained by adding a repulsive phenomenological three-body force to the too attractive two-body NNLO$_{opt}$   \cite{Knapp15}.

We are now exploring the possibility of using the chiral NNLO$_{sat}$ \cite{Ekstr15}, which includes explicitly the three-body contribution and improves the description of binding energies and nuclear radii as well \cite{Lapoux16}. Preliminary calculations using such a potential in a harmonic oscillator space encompassing up to twelve major shells yield very similar proton and neutron HF spectra for $^{16}$O and  
more compact level schemes for $^{16}$O and  $^{22}$O. In fact, the gaps between the $(sd)$ and $(0p)$ states is $\sim$8 MeV for both protons and neutrons in $^{16}$O and $\sim$ 13 MeV for  protons and $\sim$ 11 MeV for neutrons in $^{22}$O, much smaller than the corresponding gaps produced by NNLO$_{opt}$, $\sim$14 MeV in $^{16}$O and $\sim$20 MeV   in $^{22}$O. We feel therefore encouraged to pursue along this direction.

\begin{acknowledgments}
This work was partly supported by the Czech Science Foundation (Czech Republic), 16-16772S. Two of the authors (F. Knapp and P. Vesel\'y) thank the INFN (Italy) for financial support. Highly appreciated was the access to computing and storage facilities provided by the Meta Centrum
under the program LM2010005 and the CERIT-SC under the program Centre CERIT Scientific Cloud, part of the Operational Program Research and Development for Innovations, Reg. No. CZ.1.05/3.2.00/08.0144.
\end{acknowledgments}

\bibliographystyle{apsrev}
\bibliography{EMPMA1521}

\begin{thebibliography}{53}
\expandafter\ifx\csname natexlab\endcsname\relax\def\natexlab#1{#1}\fi
\expandafter\ifx\csname bibnamefont\endcsname\relax
  \def\bibnamefont#1{#1}\fi
\expandafter\ifx\csname bibfnamefont\endcsname\relax
  \def\bibfnamefont#1{#1}\fi
\expandafter\ifx\csname citenamefont\endcsname\relax
  \def\citenamefont#1{#1}\fi
\expandafter\ifx\csname url\endcsname\relax
  \def\url#1{\texttt{#1}}\fi
\expandafter\ifx\csname urlprefix\endcsname\relax\def\urlprefix{URL }\fi
\providecommand{\bibinfo}[2]{#2}
\providecommand{\eprint}[2][]{\url{#2}}

\bibitem[{\citenamefont{Mizuyama et~al.}(2012)\citenamefont{Mizuyama,
  Col$\grave{\rm{o}}$, and Vigezzi}}]{Mizu12}
\bibinfo{author}{\bibfnamefont{K.}~\bibnamefont{Mizuyama}},
  \bibinfo{author}{\bibfnamefont{G.}~\bibnamefont{Col$\grave{\rm{o}}$}},
  \bibnamefont{and} \bibinfo{author}{\bibfnamefont{E.}~\bibnamefont{Vigezzi}},
  \bibinfo{journal}{Phys Rev. C} \textbf{\bibinfo{volume}{86}},
  \bibinfo{pages}{034318} (\bibinfo{year}{2012}).

\bibitem[{\citenamefont{Cao et~al.}(2014)\citenamefont{Cao, Col\`o, Sagawa, and
  Bortignon}}]{Cao14}
\bibinfo{author}{\bibfnamefont{L.-G.} \bibnamefont{Cao}},
  \bibinfo{author}{\bibfnamefont{G.}~\bibnamefont{Col\`o}},
  \bibinfo{author}{\bibfnamefont{H.}~\bibnamefont{Sagawa}}, \bibnamefont{and}
  \bibinfo{author}{\bibfnamefont{P.~F.} \bibnamefont{Bortignon}},
  \bibinfo{journal}{Phys. Rev. C} \textbf{\bibinfo{volume}{89}},
  \bibinfo{pages}{044314} (\bibinfo{year}{2014}).

\bibitem[{\citenamefont{Tarpanov et~al.}(2014)\citenamefont{Tarpanov,
  Dobaczewski, Toivanen, and Carlsson}}]{Tarpa14}
\bibinfo{author}{\bibfnamefont{D.}~\bibnamefont{Tarpanov}},
  \bibinfo{author}{\bibfnamefont{J.}~\bibnamefont{Dobaczewski}},
  \bibinfo{author}{\bibfnamefont{J.}~\bibnamefont{Toivanen}}, \bibnamefont{and}
  \bibinfo{author}{\bibfnamefont{B.~G.} \bibnamefont{Carlsson}},
  \bibinfo{journal}{Phys. Rev. Lett.} \textbf{\bibinfo{volume}{113}},
  \bibinfo{pages}{252501} (\bibinfo{year}{2014}).

\bibitem[{\citenamefont{Afanasjev and Litvinova}(2015)}]{AfaLitv15}
\bibinfo{author}{\bibfnamefont{A.~V.} \bibnamefont{Afanasjev}}
  \bibnamefont{and}
  \bibinfo{author}{\bibfnamefont{E.}~\bibnamefont{Litvinova}},
  \bibinfo{journal}{Phys Rev. C} \textbf{\bibinfo{volume}{92}},
  \bibinfo{pages}{044317} (\bibinfo{year}{2015}).

\bibitem[{\citenamefont{Gnezdilov et~al.}(2014)\citenamefont{Gnezdilov, Borzov,
  Saperstein, and Tolokonnikov}}]{Gnez14}
\bibinfo{author}{\bibfnamefont{N.~V.} \bibnamefont{Gnezdilov}},
  \bibinfo{author}{\bibfnamefont{I.~N.} \bibnamefont{Borzov}},
  \bibinfo{author}{\bibfnamefont{E.~E.} \bibnamefont{Saperstein}},
  \bibnamefont{and} \bibinfo{author}{\bibfnamefont{S.~V.}
  \bibnamefont{Tolokonnikov}}, \bibinfo{journal}{Phys. Rev. C}
  \textbf{\bibinfo{volume}{89}}, \bibinfo{pages}{034304}
  (\bibinfo{year}{2014}).

\bibitem[{\citenamefont{Mishev and Voronov}(2008)}]{MishVor08}
\bibinfo{author}{\bibfnamefont{S.}~\bibnamefont{Mishev}} \bibnamefont{and}
  \bibinfo{author}{\bibfnamefont{V.~V.} \bibnamefont{Voronov}},
  \bibinfo{journal}{Phys Rev. C} \textbf{\bibinfo{volume}{78}},
  \bibinfo{pages}{024310} (\bibinfo{year}{2008}).

\bibitem[{\citenamefont{Co' et~al.}(2015)\citenamefont{Co', De~Donno, Anguiano,
  Bernard, and Lallena}}]{Co15}
\bibinfo{author}{\bibfnamefont{G.}~\bibnamefont{Co'}},
  \bibinfo{author}{\bibfnamefont{V.}~\bibnamefont{De~Donno}},
  \bibinfo{author}{\bibfnamefont{M.}~\bibnamefont{Anguiano}},
  \bibinfo{author}{\bibfnamefont{R.~N.} \bibnamefont{Bernard}},
  \bibnamefont{and} \bibinfo{author}{\bibfnamefont{A.~M.}
  \bibnamefont{Lallena}}, \bibinfo{journal}{Phys Rev. C}
  \textbf{\bibinfo{volume}{92}}, \bibinfo{pages}{024314}
  (\bibinfo{year}{2015}).

\bibitem[{\citenamefont{Nomura et~al.}(2017)\citenamefont{Nomura,
  Rodr\'{\i}guez-Guzm\'an, and Robledo}}]{Nomura17}
\bibinfo{author}{\bibfnamefont{K.}~\bibnamefont{Nomura}},
  \bibinfo{author}{\bibfnamefont{R.}~\bibnamefont{Rodr\'{\i}guez-Guzm\'an}},
  \bibnamefont{and} \bibinfo{author}{\bibfnamefont{L.~M.}
  \bibnamefont{Robledo}}, \bibinfo{journal}{Phys. Rev. C}
  \textbf{\bibinfo{volume}{96}}, \bibinfo{pages}{064316}
  (\bibinfo{year}{2017}).

\bibitem[{\citenamefont{Gour et~al.}(2006)\citenamefont{Gour, Piecuch,
  Hjorth-Jensen, Wloch, and Dean}}]{Gour06}
\bibinfo{author}{\bibfnamefont{J.~R.} \bibnamefont{Gour}},
  \bibinfo{author}{\bibfnamefont{P.}~\bibnamefont{Piecuch}},
  \bibinfo{author}{\bibfnamefont{M.}~\bibnamefont{Hjorth-Jensen}},
  \bibinfo{author}{\bibfnamefont{M.}~\bibnamefont{Wloch}}, \bibnamefont{and}
  \bibinfo{author}{\bibfnamefont{D.~J.} \bibnamefont{Dean}},
  \bibinfo{journal}{Phys Rev. C} \textbf{\bibinfo{volume}{74}},
  \bibinfo{pages}{024310} (\bibinfo{year}{2006}).

\bibitem[{\citenamefont{Hagen et~al.}(2010)\citenamefont{Hagen, Papenbrock, and
  Hjorth-Jensen}}]{Hagen10}
\bibinfo{author}{\bibfnamefont{G.}~\bibnamefont{Hagen}},
  \bibinfo{author}{\bibfnamefont{T.}~\bibnamefont{Papenbrock}},
  \bibnamefont{and}
  \bibinfo{author}{\bibfnamefont{M.}~\bibnamefont{Hjorth-Jensen}},
  \bibinfo{journal}{Phys. Rev. Lett.} \textbf{\bibinfo{volume}{104}},
  \bibinfo{pages}{182501} (\bibinfo{year}{2010}).

\bibitem[{\citenamefont{Hagen et~al.}(2012)\citenamefont{Hagen, Hjorth-Jensen,
  Jansen, Machleidt, and Papenbrock}}]{Hagen12}
\bibinfo{author}{\bibfnamefont{G.}~\bibnamefont{Hagen}},
  \bibinfo{author}{\bibfnamefont{M.}~\bibnamefont{Hjorth-Jensen}},
  \bibinfo{author}{\bibfnamefont{G.~R.} \bibnamefont{Jansen}},
  \bibinfo{author}{\bibfnamefont{R.}~\bibnamefont{Machleidt}},
  \bibnamefont{and}
  \bibinfo{author}{\bibfnamefont{T.}~\bibnamefont{Papenbrock}},
  \bibinfo{journal}{Phys. Rev. Lett.} \textbf{\bibinfo{volume}{108}},
  \bibinfo{pages}{242501} (\bibinfo{year}{2012}).

\bibitem[{\citenamefont{Jansen et~al.}(2014)\citenamefont{Jansen, Engel, Hagen,
  Navratil, and Signoracci}}]{Jansen14}
\bibinfo{author}{\bibfnamefont{G.~R.} \bibnamefont{Jansen}},
  \bibinfo{author}{\bibfnamefont{J.}~\bibnamefont{Engel}},
  \bibinfo{author}{\bibfnamefont{G.}~\bibnamefont{Hagen}},
  \bibinfo{author}{\bibfnamefont{P.}~\bibnamefont{Navratil}}, \bibnamefont{and}
  \bibinfo{author}{\bibfnamefont{A.}~\bibnamefont{Signoracci}},
  \bibinfo{journal}{Phys. Rev. Lett.} \textbf{\bibinfo{volume}{113}},
  \bibinfo{pages}{142502} (\bibinfo{year}{2014}).

\bibitem[{\citenamefont{Hagen et~al.}(2014)\citenamefont{Hagen, Papenbrock,
  Hjorth-Jensen, and Dean}}]{Hagen14}
\bibinfo{author}{\bibfnamefont{G.}~\bibnamefont{Hagen}},
  \bibinfo{author}{\bibfnamefont{T.}~\bibnamefont{Papenbrock}},
  \bibinfo{author}{\bibfnamefont{M.}~\bibnamefont{Hjorth-Jensen}},
  \bibnamefont{and} \bibinfo{author}{\bibfnamefont{D.~J.} \bibnamefont{Dean}},
  \bibinfo{journal}{Rep. Prog. Phys.} \textbf{\bibinfo{volume}{77}},
  \bibinfo{pages}{096302} (\bibinfo{year}{2014}).

\bibitem[{\citenamefont{Cipollone et~al.}(2013)\citenamefont{Cipollone,
  Barbieri, , and Navr\'{a}til}}]{Cipol13}
\bibinfo{author}{\bibfnamefont{A.}~\bibnamefont{Cipollone}},
  \bibinfo{author}{\bibfnamefont{C.}~\bibnamefont{Barbieri}}, ,
  \bibnamefont{and}
  \bibinfo{author}{\bibfnamefont{P.}~\bibnamefont{Navr\'{a}til}},
  \bibinfo{journal}{Phys. Rev. Lett.} \textbf{\bibinfo{volume}{111}},
  \bibinfo{pages}{062501} (\bibinfo{year}{2013}).

\bibitem[{\citenamefont{Barrett et~al.}(2013)\citenamefont{Barrett, Navr\'atil,
  and Vary}}]{Barrett13}
\bibinfo{author}{\bibfnamefont{B.~R.} \bibnamefont{Barrett}},
  \bibinfo{author}{\bibfnamefont{P.}~\bibnamefont{Navr\'atil}},
  \bibnamefont{and} \bibinfo{author}{\bibfnamefont{J.~P.} \bibnamefont{Vary}},
  \bibinfo{journal}{Progress in Particle and Nuclear Physics}
  \textbf{\bibinfo{volume}{69}}, \bibinfo{pages}{131} (\bibinfo{year}{2013}).

\bibitem[{\citenamefont{Holt et~al.}(2013)\citenamefont{Holt, Men\'endez, and
  Schwenk}}]{Holt13}
\bibinfo{author}{\bibfnamefont{J.~D.} \bibnamefont{Holt}},
  \bibinfo{author}{\bibfnamefont{J.}~\bibnamefont{Men\'endez}},
  \bibnamefont{and} \bibinfo{author}{\bibfnamefont{A.}~\bibnamefont{Schwenk}},
  \bibinfo{journal}{Eur. Phys. J. A} \textbf{\bibinfo{volume}{49}},
  \bibinfo{pages}{39} (\bibinfo{year}{2013}).

\bibitem[{\citenamefont{Andreozzi et~al.}(2007)\citenamefont{Andreozzi, Knapp,
  Lo~Iudice, Porrino, and Kvasil}}]{AndLo}
\bibinfo{author}{\bibfnamefont{F.}~\bibnamefont{Andreozzi}},
  \bibinfo{author}{\bibfnamefont{F.}~\bibnamefont{Knapp}},
  \bibinfo{author}{\bibfnamefont{N.}~\bibnamefont{Lo~Iudice}},
  \bibinfo{author}{\bibfnamefont{A.}~\bibnamefont{Porrino}}, \bibnamefont{and}
  \bibinfo{author}{\bibfnamefont{J.}~\bibnamefont{Kvasil}},
  \bibinfo{journal}{Phys. Rev. C} \textbf{\bibinfo{volume}{75}},
  \bibinfo{pages}{044312} (\bibinfo{year}{2007}).

\bibitem[{\citenamefont{Andreozzi et~al.}(2008)\citenamefont{Andreozzi, Knapp,
  Lo~Iudice, Porrino, and Kvasil}}]{AndLo1}
\bibinfo{author}{\bibfnamefont{F.}~\bibnamefont{Andreozzi}},
  \bibinfo{author}{\bibfnamefont{F.}~\bibnamefont{Knapp}},
  \bibinfo{author}{\bibfnamefont{N.}~\bibnamefont{Lo~Iudice}},
  \bibinfo{author}{\bibfnamefont{A.}~\bibnamefont{Porrino}}, \bibnamefont{and}
  \bibinfo{author}{\bibfnamefont{J.}~\bibnamefont{Kvasil}},
  \bibinfo{journal}{Phys. Rev. C} \textbf{\bibinfo{volume}{78}},
  \bibinfo{pages}{054308} (\bibinfo{year}{2008}).

\bibitem[{\citenamefont{Bianco et~al.}(2012{\natexlab{a}})\citenamefont{Bianco,
  Knapp, Lo~Iudice, Andreozzi, and Porrino}}]{Bianco}
\bibinfo{author}{\bibfnamefont{D.}~\bibnamefont{Bianco}},
  \bibinfo{author}{\bibfnamefont{F.}~\bibnamefont{Knapp}},
  \bibinfo{author}{\bibfnamefont{N.}~\bibnamefont{Lo~Iudice}},
  \bibinfo{author}{\bibfnamefont{F.}~\bibnamefont{Andreozzi}},
  \bibnamefont{and} \bibinfo{author}{\bibfnamefont{A.}~\bibnamefont{Porrino}},
  \bibinfo{journal}{Phys. Rev. C} \textbf{\bibinfo{volume}{85}},
  \bibinfo{pages}{014313} (\bibinfo{year}{2012}{\natexlab{a}}).

\bibitem[{\citenamefont{Bianco et~al.}(2012{\natexlab{b}})\citenamefont{Bianco,
  Knapp, Lo~Iudice, Andreozzi, Porrino, and Vesel\'y}}]{bianco12}
\bibinfo{author}{\bibfnamefont{D.}~\bibnamefont{Bianco}},
  \bibinfo{author}{\bibfnamefont{F.}~\bibnamefont{Knapp}},
  \bibinfo{author}{\bibfnamefont{N.}~\bibnamefont{Lo~Iudice}},
  \bibinfo{author}{\bibfnamefont{F.}~\bibnamefont{Andreozzi}},
  \bibinfo{author}{\bibfnamefont{A.}~\bibnamefont{Porrino}}, \bibnamefont{and}
  \bibinfo{author}{\bibfnamefont{P.}~\bibnamefont{Vesel\'y}},
  \bibinfo{journal}{Phys. Rev. C} \textbf{\bibinfo{volume}{86}},
  \bibinfo{pages}{044327} (\bibinfo{year}{2012}{\natexlab{b}}).

\bibitem[{\citenamefont{Knapp et~al.}(2014)\citenamefont{Knapp, Lo~Iudice,
  Vesel\'y, Andreozzi, De~Gregorio, and Porrino}}]{Knapp14}
\bibinfo{author}{\bibfnamefont{F.}~\bibnamefont{Knapp}},
  \bibinfo{author}{\bibfnamefont{N.}~\bibnamefont{Lo~Iudice}},
  \bibinfo{author}{\bibfnamefont{P.}~\bibnamefont{Vesel\'y}},
  \bibinfo{author}{\bibfnamefont{F.}~\bibnamefont{Andreozzi}},
  \bibinfo{author}{\bibfnamefont{G.}~\bibnamefont{De~Gregorio}},
  \bibnamefont{and} \bibinfo{author}{\bibfnamefont{A.}~\bibnamefont{Porrino}},
  \bibinfo{journal}{Phys. Rev. C} \textbf{\bibinfo{volume}{90}},
  \bibinfo{pages}{014310} (\bibinfo{year}{2014}).

\bibitem[{\citenamefont{Knapp et~al.}(2015)\citenamefont{Knapp, Lo~Iudice,
  Vesel\'y, Andreozzi, De~Gregorio, and Porrino}}]{Knapp15}
\bibinfo{author}{\bibfnamefont{F.}~\bibnamefont{Knapp}},
  \bibinfo{author}{\bibfnamefont{N.}~\bibnamefont{Lo~Iudice}},
  \bibinfo{author}{\bibfnamefont{P.}~\bibnamefont{Vesel\'y}},
  \bibinfo{author}{\bibfnamefont{F.}~\bibnamefont{Andreozzi}},
  \bibinfo{author}{\bibfnamefont{G.}~\bibnamefont{De~Gregorio}},
  \bibnamefont{and} \bibinfo{author}{\bibfnamefont{A.}~\bibnamefont{Porrino}},
  \bibinfo{journal}{Phys Rev. C} \textbf{\bibinfo{volume}{92}},
  \bibinfo{pages}{054315} (\bibinfo{year}{2015}).

\bibitem[{\citenamefont{De~Gregorio
  et~al.}(2016{\natexlab{a}})\citenamefont{De~Gregorio, Knapp, Lo~Iudice, and
  Vesel\'y}}]{DeGreg16}
\bibinfo{author}{\bibfnamefont{G.}~\bibnamefont{De~Gregorio}},
  \bibinfo{author}{\bibfnamefont{F.}~\bibnamefont{Knapp}},
  \bibinfo{author}{\bibfnamefont{N.}~\bibnamefont{Lo~Iudice}},
  \bibnamefont{and} \bibinfo{author}{\bibfnamefont{P.}~\bibnamefont{Vesel\'y}},
  \bibinfo{journal}{Phys Rev. C} \textbf{\bibinfo{volume}{93}},
  \bibinfo{pages}{044314} (\bibinfo{year}{2016}{\natexlab{a}}).

\bibitem[{\citenamefont{De~Gregorio
  et~al.}(2016{\natexlab{b}})\citenamefont{De~Gregorio, Knapp, Lo~Iudice, and
  Vesel\'y}}]{DeGreg16a}
\bibinfo{author}{\bibfnamefont{G.}~\bibnamefont{De~Gregorio}},
  \bibinfo{author}{\bibfnamefont{F.}~\bibnamefont{Knapp}},
  \bibinfo{author}{\bibfnamefont{N.}~\bibnamefont{Lo~Iudice}},
  \bibnamefont{and} \bibinfo{author}{\bibfnamefont{P.}~\bibnamefont{Vesel\'y}},
  \bibinfo{journal}{Phys. Rev. C} \textbf{\bibinfo{volume}{94}},
  \bibinfo{pages}{061301(R)} (\bibinfo{year}{2016}{\natexlab{b}}).

\bibitem[{\citenamefont{De~Gregorio
  et~al.}(2017{\natexlab{a}})\citenamefont{De~Gregorio, Knapp, Lo~Iudice, and
  Vesel\'y}}]{DeGreg17a}
\bibinfo{author}{\bibfnamefont{G.}~\bibnamefont{De~Gregorio}},
  \bibinfo{author}{\bibfnamefont{F.}~\bibnamefont{Knapp}},
  \bibinfo{author}{\bibfnamefont{N.}~\bibnamefont{Lo~Iudice}},
  \bibnamefont{and} \bibinfo{author}{\bibfnamefont{P.}~\bibnamefont{Vesel\'y}},
  \bibinfo{journal}{Phys. Rev. C} \textbf{\bibinfo{volume}{95}},
  \bibinfo{pages}{034327} (\bibinfo{year}{2017}{\natexlab{a}}).

\bibitem[{\citenamefont{De~Gregorio
  et~al.}(2017{\natexlab{b}})\citenamefont{De~Gregorio, Knapp, Lo~Iudice, and
  Vesel\'y}}]{DeGreg17b}
\bibinfo{author}{\bibfnamefont{G.}~\bibnamefont{De~Gregorio}},
  \bibinfo{author}{\bibfnamefont{F.}~\bibnamefont{Knapp}},
  \bibinfo{author}{\bibfnamefont{N.}~\bibnamefont{Lo~Iudice}},
  \bibnamefont{and} \bibinfo{author}{\bibfnamefont{P.}~\bibnamefont{Vesel\'y}},
  \bibinfo{journal}{Phys. Scr.} \textbf{\bibinfo{volume}{92}},
  \bibinfo{pages}{074003} (\bibinfo{year}{2017}{\natexlab{b}}).

\bibitem[{\citenamefont{De~Gregorio et~al.}(2018)\citenamefont{De~Gregorio,
  Knapp, Lo~Iudice, and Vesel\'y}}]{DeGreg18}
\bibinfo{author}{\bibfnamefont{G.}~\bibnamefont{De~Gregorio}},
  \bibinfo{author}{\bibfnamefont{F.}~\bibnamefont{Knapp}},
  \bibinfo{author}{\bibfnamefont{N.}~\bibnamefont{Lo~Iudice}},
  \bibnamefont{and} \bibinfo{author}{\bibfnamefont{P.}~\bibnamefont{Vesel\'y}},
  \bibinfo{journal}{Phys. Rev. C} \textbf{\bibinfo{volume}{97}},
  \bibinfo{pages}{034311} (\bibinfo{year}{2018}).

\bibitem[{\citenamefont{Brown and Green}(1966)}]{BrownGreen66}
\bibinfo{author}{\bibfnamefont{G.~E.} \bibnamefont{Brown}} \bibnamefont{and}
  \bibinfo{author}{\bibfnamefont{A.~M.} \bibnamefont{Green}},
  \bibinfo{journal}{Nucl. Phys.} \textbf{\bibinfo{volume}{75}},
  \bibinfo{pages}{401} (\bibinfo{year}{1966}).

\bibitem[{\citenamefont{Halbert and French}(1957)}]{HalFrench57}
\bibinfo{author}{\bibfnamefont{E.~C.} \bibnamefont{Halbert}} \bibnamefont{and}
  \bibinfo{author}{\bibfnamefont{J.~B.} \bibnamefont{French}},
  \bibinfo{journal}{Phys. Rev.} \textbf{\bibinfo{volume}{105}},
  \bibinfo{pages}{1563} (\bibinfo{year}{1957}).

\bibitem[{\citenamefont{Shukla and Brown}(1968)}]{Shukla68}
\bibinfo{author}{\bibfnamefont{A.~P.} \bibnamefont{Shukla}} \bibnamefont{and}
  \bibinfo{author}{\bibfnamefont{G.~E.} \bibnamefont{Brown}},
  \bibinfo{journal}{Nucl. Phys. A} \textbf{\bibinfo{volume}{112}},
  \bibinfo{pages}{296} (\bibinfo{year}{1968}).

\bibitem[{\citenamefont{Lie et~al.}(1970)\citenamefont{Lie, Engeland, and
  Dhall}}]{Lie70}
\bibinfo{author}{\bibfnamefont{S.}~\bibnamefont{Lie}},
  \bibinfo{author}{\bibfnamefont{T.}~\bibnamefont{Engeland}}, \bibnamefont{and}
  \bibinfo{author}{\bibfnamefont{G.}~\bibnamefont{Dhall}},
  \bibinfo{journal}{Nucl. Phys. A} \textbf{\bibinfo{volume}{156}},
  \bibinfo{pages}{449} (\bibinfo{year}{1970}).

\bibitem[{\citenamefont{Alburger and Millener}(1979)}]{Alburger79}
\bibinfo{author}{\bibfnamefont{D.~E.} \bibnamefont{Alburger}} \bibnamefont{and}
  \bibinfo{author}{\bibfnamefont{D.~J.} \bibnamefont{Millener}},
  \bibinfo{journal}{Phys. Rev. C} \textbf{\bibinfo{volume}{20}},
  \bibinfo{pages}{1891} (\bibinfo{year}{1979}).

\bibitem[{\citenamefont{Raman et~al.}(1994)\citenamefont{Raman, Jurney,
  Starner, Kuronen, Keinonen, Nordlund, and Millener}}]{Raman94}
\bibinfo{author}{\bibfnamefont{S.}~\bibnamefont{Raman}},
  \bibinfo{author}{\bibfnamefont{E.~T.} \bibnamefont{Jurney}},
  \bibinfo{author}{\bibfnamefont{J.~W.} \bibnamefont{Starner}},
  \bibinfo{author}{\bibfnamefont{A.}~\bibnamefont{Kuronen}},
  \bibinfo{author}{\bibfnamefont{J.}~\bibnamefont{Keinonen}},
  \bibinfo{author}{\bibfnamefont{K.}~\bibnamefont{Nordlund}}, \bibnamefont{and}
  \bibinfo{author}{\bibfnamefont{D.~J.} \bibnamefont{Millener}},
  \bibinfo{journal}{Phys. Rev. C} \textbf{\bibinfo{volume}{50}},
  \bibinfo{pages}{682} (\bibinfo{year}{1994}).

\bibitem[{\citenamefont{Lie and Engeland}(1976)}]{LieEng76}
\bibinfo{author}{\bibfnamefont{S.}~\bibnamefont{Lie}} \bibnamefont{and}
  \bibinfo{author}{\bibfnamefont{T.}~\bibnamefont{Engeland}},
  \bibinfo{journal}{Nucl. Phys. A} \textbf{\bibinfo{volume}{267}},
  \bibinfo{pages}{123} (\bibinfo{year}{1976}).

\bibitem[{\citenamefont{Utsuno and Chiba}(2011)}]{Utsuno2011}
\bibinfo{author}{\bibfnamefont{Y.}~\bibnamefont{Utsuno}} \bibnamefont{and}
  \bibinfo{author}{\bibfnamefont{S.}~\bibnamefont{Chiba}},
  \bibinfo{journal}{Phys. Rev. C} \textbf{\bibinfo{volume}{83}},
  \bibinfo{pages}{021301} (\bibinfo{year}{2011}).

\bibitem[{\citenamefont{Mertin et~al.}(2015)\citenamefont{Mertin, Caussyn,
  Keeley, Kemper, Momotyuk, Roeder, and Volya}}]{Mertin2015}
\bibinfo{author}{\bibfnamefont{C.~E.} \bibnamefont{Mertin}},
  \bibinfo{author}{\bibfnamefont{A.~M.} \bibnamefont{Caussyn},
  \bibfnamefont{D.~D. and.~Crisp}},
  \bibinfo{author}{\bibfnamefont{N.}~\bibnamefont{Keeley}},
  \bibinfo{author}{\bibfnamefont{K.~W.} \bibnamefont{Kemper}},
  \bibinfo{author}{\bibfnamefont{O.}~\bibnamefont{Momotyuk}},
  \bibinfo{author}{\bibfnamefont{B.~T.} \bibnamefont{Roeder}},
  \bibnamefont{and} \bibinfo{author}{\bibfnamefont{A.}~\bibnamefont{Volya}},
  \bibinfo{journal}{Phys. Rev. C} \textbf{\bibinfo{volume}{91}},
  \bibinfo{pages}{044317} (\bibinfo{year}{2015}).

\bibitem[{\citenamefont{Ma et~al.}(2016)\citenamefont{Ma, Dong, Yan, Zhang,
  Yuan, Zhu, and Zhang}}]{Ma2016}
\bibinfo{author}{\bibfnamefont{H.-L.} \bibnamefont{Ma}},
  \bibinfo{author}{\bibfnamefont{B.-G.} \bibnamefont{Dong}},
  \bibinfo{author}{\bibfnamefont{Y.-L.} \bibnamefont{Yan}},
  \bibinfo{author}{\bibfnamefont{H.-Q.} \bibnamefont{Zhang}},
  \bibinfo{author}{\bibfnamefont{D.-Q.} \bibnamefont{Yuan}},
  \bibinfo{author}{\bibfnamefont{S.-Y.} \bibnamefont{Zhu}}, \bibnamefont{and}
  \bibinfo{author}{\bibfnamefont{X.-Z.} \bibnamefont{Zhang}},
  \bibinfo{journal}{Phys. Rev. C} \textbf{\bibinfo{volume}{93}},
  \bibinfo{pages}{014317} (\bibinfo{year}{2016}).

\bibitem[{\citenamefont{Catford et~al.}(1989)\citenamefont{Catford, Fifield,
  Orr, and Woods}}]{Catford89}
\bibinfo{author}{\bibfnamefont{W.}~\bibnamefont{Catford}},
  \bibinfo{author}{\bibfnamefont{L.}~\bibnamefont{Fifield}},
  \bibinfo{author}{\bibfnamefont{N.}~\bibnamefont{Orr}}, \bibnamefont{and}
  \bibinfo{author}{\bibfnamefont{C.}~\bibnamefont{Woods}},
  \bibinfo{journal}{Nucl. Phys. A} \textbf{\bibinfo{volume}{503}},
  \bibinfo{pages}{263 } (\bibinfo{year}{1989}).

\bibitem[{\citenamefont{Sauvan et~al.}(2000)\citenamefont{Sauvan, Carstoiu,
  Orr, Ang\'elique, Catford, Clarke, Cormick, Curtis, Freer, Gr\'evy
  et~al.}}]{Sauvan00}
\bibinfo{author}{\bibfnamefont{E.}~\bibnamefont{Sauvan}},
  \bibinfo{author}{\bibfnamefont{F.}~\bibnamefont{Carstoiu}},
  \bibinfo{author}{\bibfnamefont{N.}~\bibnamefont{Orr}},
  \bibinfo{author}{\bibfnamefont{J.}~\bibnamefont{Ang\'elique}},
  \bibinfo{author}{\bibfnamefont{W.}~\bibnamefont{Catford}},
  \bibinfo{author}{\bibfnamefont{N.}~\bibnamefont{Clarke}},
  \bibinfo{author}{\bibfnamefont{M.~M.} \bibnamefont{Cormick}},
  \bibinfo{author}{\bibfnamefont{N.}~\bibnamefont{Curtis}},
  \bibinfo{author}{\bibfnamefont{M.}~\bibnamefont{Freer}},
  \bibinfo{author}{\bibfnamefont{S.}~\bibnamefont{Gr\'evy}},
  \bibnamefont{et~al.}, \bibinfo{journal}{Physics Letters B}
  \textbf{\bibinfo{volume}{491}}, \bibinfo{pages}{1 } (\bibinfo{year}{2000}).

\bibitem[{\citenamefont{Sauvan et~al.}(2004)\citenamefont{Sauvan, Carstoiu,
  Orr, Winfield, Freer, Ang\'elique, Catford, Clarke, Curtis, Gr\'evy
  et~al.}}]{Sauvan04}
\bibinfo{author}{\bibfnamefont{E.}~\bibnamefont{Sauvan}},
  \bibinfo{author}{\bibfnamefont{F.}~\bibnamefont{Carstoiu}},
  \bibinfo{author}{\bibfnamefont{N.~A.} \bibnamefont{Orr}},
  \bibinfo{author}{\bibfnamefont{J.~S.} \bibnamefont{Winfield}},
  \bibinfo{author}{\bibfnamefont{M.}~\bibnamefont{Freer}},
  \bibinfo{author}{\bibfnamefont{J.~C.} \bibnamefont{Ang\'elique}},
  \bibinfo{author}{\bibfnamefont{W.~N.} \bibnamefont{Catford}},
  \bibinfo{author}{\bibfnamefont{N.~M.} \bibnamefont{Clarke}},
  \bibinfo{author}{\bibfnamefont{N.}~\bibnamefont{Curtis}},
  \bibinfo{author}{\bibfnamefont{S.}~\bibnamefont{Gr\'evy}},
  \bibnamefont{et~al.}, \bibinfo{journal}{Phys. Rev. C}
  \textbf{\bibinfo{volume}{69}}, \bibinfo{pages}{044603}
  (\bibinfo{year}{2004}).

\bibitem[{\citenamefont{Stanoiu et~al.}(2004)\citenamefont{Stanoiu, Azaiez,
  Dombr\'adi, Sorlin, Brown, Belleguic, Sohler, Saint~Laurent, Lopez-Jimenez,
  Penionzhkevich et~al.}}]{Stanoiu04}
\bibinfo{author}{\bibfnamefont{M.}~\bibnamefont{Stanoiu}},
  \bibinfo{author}{\bibfnamefont{F.}~\bibnamefont{Azaiez}},
  \bibinfo{author}{\bibfnamefont{Z.}~\bibnamefont{Dombr\'adi}},
  \bibinfo{author}{\bibfnamefont{O.}~\bibnamefont{Sorlin}},
  \bibinfo{author}{\bibfnamefont{B.~A.} \bibnamefont{Brown}},
  \bibinfo{author}{\bibfnamefont{M.}~\bibnamefont{Belleguic}},
  \bibinfo{author}{\bibfnamefont{D.}~\bibnamefont{Sohler}},
  \bibinfo{author}{\bibfnamefont{M.~G.} \bibnamefont{Saint~Laurent}},
  \bibinfo{author}{\bibfnamefont{M.~J.} \bibnamefont{Lopez-Jimenez}},
  \bibinfo{author}{\bibfnamefont{Y.~E.} \bibnamefont{Penionzhkevich}},
  \bibnamefont{et~al.}, \bibinfo{journal}{Phys. Rev. C}
  \textbf{\bibinfo{volume}{69}}, \bibinfo{pages}{034312}
  (\bibinfo{year}{2004}).

\bibitem[{\citenamefont{Mueller et~al.}(1990)\citenamefont{Mueller,
  Guillemaud-Mueller, Jacmart, Kashy, Pougheon, Richard, Staudt,
  Klapdor-Kleingrothaus, Lewitowicz, Anne et~al.}}]{Mueller90}
\bibinfo{author}{\bibfnamefont{A.}~\bibnamefont{Mueller}},
  \bibinfo{author}{\bibfnamefont{D.}~\bibnamefont{Guillemaud-Mueller}},
  \bibinfo{author}{\bibfnamefont{J.}~\bibnamefont{Jacmart}},
  \bibinfo{author}{\bibfnamefont{E.}~\bibnamefont{Kashy}},
  \bibinfo{author}{\bibfnamefont{F.}~\bibnamefont{Pougheon}},
  \bibinfo{author}{\bibfnamefont{A.}~\bibnamefont{Richard}},
  \bibinfo{author}{\bibfnamefont{A.}~\bibnamefont{Staudt}},
  \bibinfo{author}{\bibfnamefont{H.}~\bibnamefont{Klapdor-Kleingrothaus}},
  \bibinfo{author}{\bibfnamefont{M.}~\bibnamefont{Lewitowicz}},
  \bibinfo{author}{\bibfnamefont{R.}~\bibnamefont{Anne}}, \bibnamefont{et~al.},
  \bibinfo{journal}{Nucl. Phys. A} \textbf{\bibinfo{volume}{513}},
  \bibinfo{pages}{1 } (\bibinfo{year}{1990}).

\bibitem[{\citenamefont{Li et~al.}(2009)\citenamefont{Li, Lou, Ye, Hua, Jiang,
  Li, Zhang, Zheng, Ge, Kong et~al.}}]{Li09}
\bibinfo{author}{\bibfnamefont{Z.~H.} \bibnamefont{Li}},
  \bibinfo{author}{\bibfnamefont{J.~L.} \bibnamefont{Lou}},
  \bibinfo{author}{\bibfnamefont{Y.~L.} \bibnamefont{Ye}},
  \bibinfo{author}{\bibfnamefont{H.}~\bibnamefont{Hua}},
  \bibinfo{author}{\bibfnamefont{D.~X.} \bibnamefont{Jiang}},
  \bibinfo{author}{\bibfnamefont{X.~Q.} \bibnamefont{Li}},
  \bibinfo{author}{\bibfnamefont{S.~Q.} \bibnamefont{Zhang}},
  \bibinfo{author}{\bibfnamefont{T.}~\bibnamefont{Zheng}},
  \bibinfo{author}{\bibfnamefont{Y.~C.} \bibnamefont{Ge}},
  \bibinfo{author}{\bibfnamefont{Z.}~\bibnamefont{Kong}}, \bibnamefont{et~al.},
  \bibinfo{journal}{Phys. Rev. C} \textbf{\bibinfo{volume}{80}},
  \bibinfo{pages}{054315} (\bibinfo{year}{2009}).

\bibitem[{\citenamefont{Zhang}(2016)}]{Zhang16}
\bibinfo{author}{\bibfnamefont{Y.-M.} \bibnamefont{Zhang}},
  \bibinfo{journal}{J. Phys. G: Nucl. Part. Phys.}
  \textbf{\bibinfo{volume}{43}}, \bibinfo{pages}{045104}
  (\bibinfo{year}{2016}).

\bibitem[{\citenamefont{Ekstr\"om et~al.}(2013)\citenamefont{Ekstr\"om,
  Baardsen, Forss\'en, Hagen, Hjorth-Jensen, Jansen, Machleidt, Nazarewicz,
  Papenbrock, Sarich et~al.}}]{Ekstr13}
\bibinfo{author}{\bibfnamefont{A.}~\bibnamefont{Ekstr\"om}},
  \bibinfo{author}{\bibfnamefont{G.}~\bibnamefont{Baardsen}},
  \bibinfo{author}{\bibfnamefont{C.}~\bibnamefont{Forss\'en}},
  \bibinfo{author}{\bibfnamefont{G.}~\bibnamefont{Hagen}},
  \bibinfo{author}{\bibfnamefont{M.}~\bibnamefont{Hjorth-Jensen}},
  \bibinfo{author}{\bibfnamefont{G.~R.} \bibnamefont{Jansen}},
  \bibinfo{author}{\bibfnamefont{R.}~\bibnamefont{Machleidt}},
  \bibinfo{author}{\bibfnamefont{W.}~\bibnamefont{Nazarewicz}},
  \bibinfo{author}{\bibfnamefont{T.}~\bibnamefont{Papenbrock}},
  \bibinfo{author}{\bibfnamefont{J.}~\bibnamefont{Sarich}},
  \bibnamefont{et~al.}, \bibinfo{journal}{Phys. Rev. Lett.}
  \textbf{\bibinfo{volume}{110}}, \bibinfo{pages}{192502}
  (\bibinfo{year}{2013}).

\bibitem[{\citenamefont{Ajzenberg-Selove}(1991)}]{AjzeSelo91}
\bibinfo{author}{\bibfnamefont{F.}~\bibnamefont{Ajzenberg-Selove}},
  \bibinfo{journal}{Nucl. Phys. A} \textbf{\bibinfo{volume}{523}},
  \bibinfo{pages}{1 } (\bibinfo{year}{1991}).

\bibitem[{\citenamefont{Bianco et~al.}(2014)\citenamefont{Bianco, Knapp,
  Lo~Iudice, Vesel\'y, Andreozzi, De~Gregorio, and Porrino}}]{Bianco14}
\bibinfo{author}{\bibfnamefont{D.}~\bibnamefont{Bianco}},
  \bibinfo{author}{\bibfnamefont{F.}~\bibnamefont{Knapp}},
  \bibinfo{author}{\bibfnamefont{N.}~\bibnamefont{Lo~Iudice}},
  \bibinfo{author}{\bibfnamefont{P.}~\bibnamefont{Vesel\'y}},
  \bibinfo{author}{\bibfnamefont{F.}~\bibnamefont{Andreozzi}},
  \bibinfo{author}{\bibfnamefont{G.}~\bibnamefont{De~Gregorio}},
  \bibnamefont{and} \bibinfo{author}{\bibfnamefont{A.}~\bibnamefont{Porrino}},
  \bibinfo{journal}{J. Phys. G: Nucl. Part. Phys.}
  \textbf{\bibinfo{volume}{41}}, \bibinfo{pages}{025109}
  (\bibinfo{year}{2014}).

\bibitem[{\citenamefont{Bates et~al.}(1989)\citenamefont{Bates, Rassool, Milne,
  Thompson, and McNeill}}]{Bates89}
\bibinfo{author}{\bibfnamefont{A.~D.} \bibnamefont{Bates}},
  \bibinfo{author}{\bibfnamefont{R.~P.} \bibnamefont{Rassool}},
  \bibinfo{author}{\bibfnamefont{E.~A.} \bibnamefont{Milne}},
  \bibinfo{author}{\bibfnamefont{M.~N.} \bibnamefont{Thompson}},
  \bibnamefont{and} \bibinfo{author}{\bibfnamefont{K.~G.}
  \bibnamefont{McNeill}}, \bibinfo{journal}{Phys. Rev. C}
  \textbf{\bibinfo{volume}{40}}, \bibinfo{pages}{506} (\bibinfo{year}{1989}).

\bibitem[{\citenamefont{Firestone}(2015)}]{Firestone15}
\bibinfo{author}{\bibfnamefont{R.}~\bibnamefont{Firestone}},
  \bibinfo{journal}{Nuclear Data Sheets} \textbf{\bibinfo{volume}{127}},
  \bibinfo{pages}{1 } (\bibinfo{year}{2015}).

\bibitem[{\citenamefont{Haxton and Johnson}(1990)}]{HaxJoh90}
\bibinfo{author}{\bibfnamefont{W.~C.} \bibnamefont{Haxton}} \bibnamefont{and}
  \bibinfo{author}{\bibfnamefont{C.}~\bibnamefont{Johnson}},
  \bibinfo{journal}{Phys. Rev. Lett.} \textbf{\bibinfo{volume}{65}},
  \bibinfo{pages}{1325} (\bibinfo{year}{1990}).

\bibitem[{\citenamefont{Warburton et~al.}(1992)\citenamefont{Warburton, Brown,
  and Millener}}]{WarBrowMil92}
\bibinfo{author}{\bibfnamefont{E.}~\bibnamefont{Warburton}},
  \bibinfo{author}{\bibfnamefont{B.}~\bibnamefont{Brown}}, \bibnamefont{and}
  \bibinfo{author}{\bibfnamefont{D.}~\bibnamefont{Millener}},
  \bibinfo{journal}{Phys. Lett. B} \textbf{\bibinfo{volume}{293}},
  \bibinfo{pages}{7 } (\bibinfo{year}{1992}).

\bibitem[{\citenamefont{Ekstr\"om et~al.}(2015)\citenamefont{Ekstr\"om, Jansen,
  Wendt, Hagen, Papenbrock, Carlsson, Forss\'en, Hjorth-Jensen, Navr\'atil, and
  Nazarewicz}}]{Ekstr15}
\bibinfo{author}{\bibfnamefont{A.}~\bibnamefont{Ekstr\"om}},
  \bibinfo{author}{\bibfnamefont{G.~R.} \bibnamefont{Jansen}},
  \bibinfo{author}{\bibfnamefont{K.~A.} \bibnamefont{Wendt}},
  \bibinfo{author}{\bibfnamefont{G.}~\bibnamefont{Hagen}},
  \bibinfo{author}{\bibfnamefont{T.}~\bibnamefont{Papenbrock}},
  \bibinfo{author}{\bibfnamefont{B.~D.} \bibnamefont{Carlsson}},
  \bibinfo{author}{\bibfnamefont{C.}~\bibnamefont{Forss\'en}},
  \bibinfo{author}{\bibfnamefont{M.}~\bibnamefont{Hjorth-Jensen}},
  \bibinfo{author}{\bibfnamefont{P.}~\bibnamefont{Navr\'atil}},
  \bibnamefont{and}
  \bibinfo{author}{\bibfnamefont{W.}~\bibnamefont{Nazarewicz}},
  \bibinfo{journal}{Phys. Rev. C} \textbf{\bibinfo{volume}{91}},
  \bibinfo{pages}{051301} (\bibinfo{year}{2015}).

\bibitem[{\citenamefont{Lapoux et~al.}(2016)\citenamefont{Lapoux,
  Som$\grave{\rm{a}}$, Barbieri, Hergert, Holt, and Stroberg}}]{Lapoux16}
\bibinfo{author}{\bibfnamefont{V.}~\bibnamefont{Lapoux}},
  \bibinfo{author}{\bibfnamefont{V.}~\bibnamefont{Som$\grave{\rm{a}}$}},
  \bibinfo{author}{\bibfnamefont{C.}~\bibnamefont{Barbieri}},
  \bibinfo{author}{\bibfnamefont{H.}~\bibnamefont{Hergert}},
  \bibinfo{author}{\bibfnamefont{J.~D.} \bibnamefont{Holt}}, \bibnamefont{and}
  \bibinfo{author}{\bibfnamefont{S.~R.} \bibnamefont{Stroberg}},
  \bibinfo{journal}{Phys. Rev. Lett.} \textbf{\bibinfo{volume}{117}},
  \bibinfo{pages}{052501} (\bibinfo{year}{2016}).

\end{thebibliography}
\end{document}